\documentclass[aps,prb,twocolumn,showpacs,amsmath,amssymb,reprint]{revtex4-2}
\usepackage{dcolumn}
\usepackage{graphicx}
\usepackage{bm}
\begin{document}
	\title{Ferromagnetic exchange field stabilized antiferromagnetic ordering in a cuprate superconductor} 
	\author{Biswajit Dutta and A. Banerjee}
	\affiliation{UGC-DAE Consortium for Scientific Research, University Campus, Khandwa Road, Indore-452001, India.}
	
	\begin{abstract}
		We report experimental evidence of formation of antiferromagnetic clusters well within cuprate superconductor $La_{1.85}Sr_{0.15}CuO_{4}$ (LCu) in a composite made of LCu and ferromagnet $La_{0.6}Sr_{0.4}CoO_{3}$ (LCo). It is found that the exchange field of LCo suppress the dynamic antiferromagnetic spin fluctuation of LCu and short range ordered superparamagnetic type antiferromagnetic (AFM) clusters are formed at the cost of superconducting volume fraction. With the help of linear and non-linear ac-susceptibility measurement we show the evidence of thermal blocking of these antiferromagnetic clusters. Further, we show that the shrank superconducting volume fractions undergo quantum size effect (QSE) and follow DeGennes-Tinkham theory on finite size effect of superconductor. These give clear indication that antiferromagnetic spin fluctuation can be a mediator of electron pairing in cuprate superconductor.
	\end{abstract}
	
\maketitle
	\section {INTRODUCTION}
	\vspace{-4.5mm}
	One of the long lived mystery associated with high temperature cuprate superconductor is the origin of coherent electron pairing in these compounds. The magnetic state of a parent cuprate is  3D Neel antiferromagnet (AFM). After doping holes or electrons the ordered antiferromagnetic state destroys and superconducting state emerges \cite{c4t2}. However, existence of AFM type spin correlation is reveled throughout the hole or electron doped phase diagram (including the doping region where superconductivity exists) and depending on dynamical nature of the AFM  correlation the strength of superconductivity is decided \cite{c4t2,c4r1,c4r2,c4t,c4t1,c4t3,c4m1}. It is also observed that when the superconducting state is destroyed by doping or by applying very high magnetic field, then emergence of AFM like state is observed. Like  Neodymium (Nd) doping in $La_{1.85}Sr_{0.15}CuO_{4}$ (LCu) destroys the superconducting state and results in antiferromagnetic type CDW phase \cite{c4r50,c4r51,c4r52}. Further, suppression of superconductivity and emergence of Charge Density Wave (CDW) phase or AFM state is also reveled in ferromagnet (FM) and cuprate superconductor heterostructures at normal condition (or without imposing any extreme condition or doping in the superconductor) \cite{c4m2,c4m3,c4r9,c4r10,c4r11,c4r13,c4r14,c4r15}. These characteristic behavior indicates a close correlation between  AFM ordering and coherent electron pairing mechanism in cuprate superconductors, which has also opened up new perspective to understand the superconducting pairing mechanism of cuprate superconductors. Therefore,  a detailed study of cuprate superconductor in proximity to FM systems are required to get further knowledge about the involvement of antiferromagnetic type spin correlation in the pairing mechanism. Mostly, such studies on oxide superconductor and ferromagnet (SC/FM) are based on  YBa$_{2}$Cu$_{3}$O$_{7}$ (YBCO) and La$_{0.7}$Sr${0.3}$MnO$_{3}$ (or La$_{0.7}$Ca${0.3}$MnO$_{3}$). However, the problem associated with these YBCO based SC/FM heterostructures is high probability of disruption of CuO chain (which act as charge reservoir in YBCO) across the interface. As a result of that the hole concentration of YBCO gets modified, changing the physical property in bulk like manner unrelated to interface physics \cite{c4r16,c4y1,c4y2,c4y3}. It has been observed that the SC/FM interfaces consisting of $La_{1.85}Sr_{0.15}CuO_{4}$ (LCu) as superconductor does not show any dominant charge transfer or orbital reconstruction phenomena to take place across the interface \cite{c4a1,c4r19,c4r17,c4r18} and maintained their charge state intact with respect to the parent ingredient. Hence, LCu based (or LCu type i.e. Pr$_{1-x}$Ce$_{x}$CuO$_4$ \cite{c4a1}) FM/SC composites (or bilayer) are ideal to study the effects of exchange field \cite{c4a1}, as well as the interfacial strain effect  on the magnetic properties of cuprate superconductors \cite{c4a1,c4r19,c4r17,c4r18}.
	
	 Here, we present a study on composites made of superconductor LCu and a ferromagnet La$_{0.6}$Sr$_{0.4}$CoO$_3$ (LCo) to explore the effect of  exchange bias field on the hole doped cuprate superconductor. LCo is chosen as a ferromagnetic counterpart because it shows long-range ferromagnetic ordering and large magneto-crystalline anisotropy \cite{c4r20,c4r21,c4r22,c4r23}, which can facilitate substantial amount of exchange bias field on the spins of copper atom of LCu. In this respect, it has an advantage over manganites (La$_{0.7}$Sr$_{0.3}$MnO$_3$, La$_{0.7}$Ca$_{0.3}$MnO$_3$ \cite{c4r19,c4r17,c4r18}) to explore the effect of exchange field on cuprate superconductors.
	We show that the dynamic antiferromagnetic spin fluctuation of LCu   \cite{c4r50,c4r51,c4r52,c4r58,c4r59,c4r60} is suppressed due to the magnetic exchange field of LCo and short range order AFM phase is developed at the cost of superconducting volume fraction. As a result, the bulk superconducting region shrank to finite size clusters, and quantum size effect (QSE) of the superconductor appears, without any reduction in the crystallite size of the superconductor. The exchange field amplitude on LCu is tuned by changing the effective interface as well as by dc magnetic field. The effective interface  is varied by following several processes like, reducing (or increasing) the particle size, decreasing (or enhancing) the concentration of LCo in the LCu matrix \cite{c4r24} and grinding the composite pallet.
	
We have used Linear and nonlinear ac-susceptibility measurements to study the proximity effect induced change of the magnetic state of a superconductor in a SC/FM interface. Nonlinear susceptibilities are found to be very effective tool in determining various characteristics of a type-2 superconductor, such as to identify the onset point of irreversible flux motion, nature of flux dynamic and also very effective in determining various critical thermodynamical parameters (like critical current, critical field and critical temperature) \cite{c4r27,c4r28,c4r29,c4r30}. This technique is also very effective in unambiguously distinguishing various metastable states like spin-glass, cluster-glass and superparamagnet \cite{c4x,c4x1,c4x2,c4x3} and also in establishing the nature of magnetic ground state (like ferromagnetic or antiferromagnetic) \cite{c4x4,c4x5}.
	
In composite (or in heterostructure) the interplay between different electronic ground states are modulated through the extended interface effect, but in case of oxide heterostructures oxygen stoichiometry plays a big role in modifying the physical property which is unrelated to interface physics. Hence it is very important to determine the oxygen stoichiometry (which defines the spin state) of the ingredients with respect to their parent compounds for a conclusive discrimination between the phenomena related to interface effect and the phenomena associated with the degradation of oxygen stoichiometry. We have proposed a method to determine the spin state of a ferromagnet and hole concentration of a superconductor in a SC/FM composite system by using the low field magnetic ac-susceptibility technique and XRD measurement, which is already published in Ref.\cite{c4r24} and briefly discussed here.
 
	 \maketitle\section{SAMPLE PREPARATION AND CHARACTERIZATION }
The parent ingredient LCu and LCo is prepared by pyrophoric method. Then the composites are prepared by solid state reaction method, preparation technique is more elaborately explained in Ref.\cite{c4r24}. The precursor of LCo just obtained after pyrophoric chemical reaction is annealed at different temperatures, viz. $950^0 C$, $900^0 C$ and $850^0 C$ to obtained three different crystalline size (CS) which are nomenclatured as LCo950, LCo900 and LCo850 respectively ((CS)$_ {LCo850}$$<$(CS)$_ {LCo900}$$<$ (CS)$_ {LCo950}$). Then 76 weight percentage of LCu and 24 weight percentage of LCo950 is mixed, palletized at comparatively high pressure (i.e., 150 kilo-Newton (kN))  and heated at 800$^{0}C$ to prepare composite A1. In the  similar way A2 and A3 is prepared by mixing LCo900 and LCo850 with LCu respectively. The concentration of LCo850 is also varied in LCu matrix to change the effective interface between LCu and LCo. In this case LCu and LCo850 is mixed at different weight ratio like 76:24, 85:15 and 95:5 and these composites are nomenclatured as A3, A4 and A5, respectively. The parent ingredients are independently grinned palletized at similar pressure (i.e., 150kN) and annealed at same temperature (i.e. 800\,$^0$C) for comparison purposes. The weight percentage detail of the corresponding parent ingredient present in the composites (i.e., A1, A2, A3, A4, A5) and  detailed nomenclature of all composites along with the crystalline size of parent LCu and LCo are provided in \texttt{TABLE\,I}. The crystalline size of the ferromagnet and the superconductor is calculated from the XRD pattern as well as from Transmission electron microscopy (TEM) measurement.
\begin{table*}[t]
		\caption{List of composites and crystalline size details}
		\setlength{\tabcolsep}{18pt}
		\renewcommand{\arraystretch}{1.5}
		\begin{center}
			\begin{tabular}{| p{1cm}| p{4cm}| p{3cm}| p{3cm}| }
				\hline
				Composite Name&LCu [Weight\%] + LCo (annealing Temperature) [Weight\%] & average crystalline Size of LCo (nm) \\  
				\hline\hline
				A1 & LCu[76\%]+LCo(950)[24\%] & 69(nm) \\
				
				A2 & LCu[76\%]+LCo(900)[24\%] & 47(nm)\\
				
				A3 & LCu[76\%]+LCo(850)[24\%] & 27(nm)\\
				
				A4 & LCu[85\%]+LCo(850)[15\%] & 27(nm)\\
				
				A5 & LCu[95\%]+LCo(850)[5\%] & 27(nm)\\
				\hline
			\end{tabular}
		\end{center}
	\end{table*}  
	
	\subsection{Structural Characterization : XRD}
	X-ray diffraction (XRD) measurements are performed in Bruker X-ray diffractometer from $10^{0}$ - $90^{0}$  at an interval of $0.02^{0}$. \texttt{FIG.\,1(a)} shows the two-phase Rietveld refinement of the composite A3 and \texttt{FIG.\,1(b)} shows the XRD pattern of A1, A2 and A3 respectively. \texttt{FIG.\,2(a)} shows the Rietveld refinement  of the composite A5 and the combined plot of the XRD spectrum of A3, A4 and A5 composite are shown in \texttt{FIG.\,2(b)} for comparison purpose. The  phase fractions and other structural details obtained from two phase Rietveld refinement are given in \texttt{TABLE\,II}.
	
	\begin{figure}[htbp]
		\centering
		\includegraphics[width=9cm]{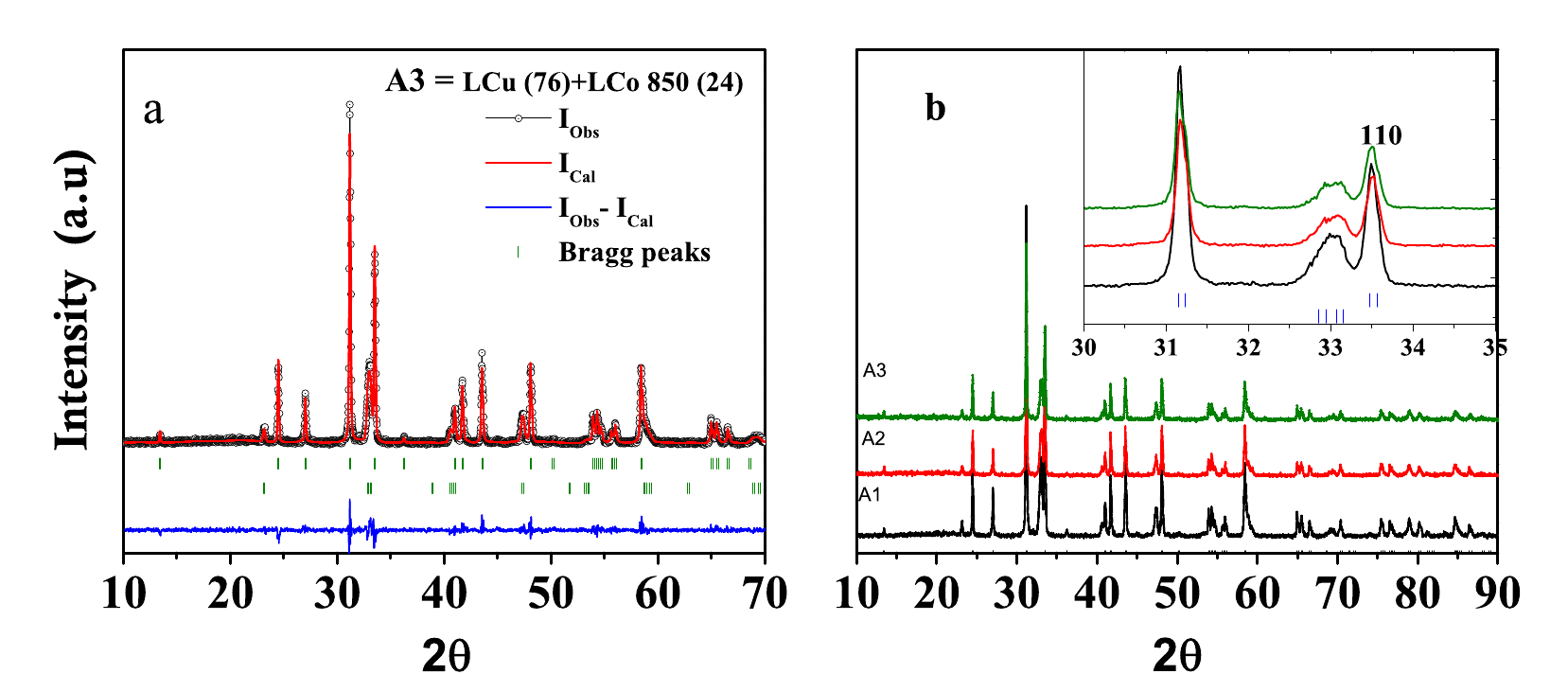}
		\vspace*{-8mm}
		\caption{(Colour Online) (a) Rietveld refinement of the composite A3. (b) XRD plot of all composites A1, A2, A3 (Inset shows the region $30^{0}$-$35^{0}$ shows the maximum intensity peak region)}
		\label{fig:fig1}
	\end{figure}
		
		\begin{table}
		\caption{Best fitted parameters, SG-Space group, PF-Phase fraction, a, b and c are lattice parameters} 
		\begin{tabular}{|c| c| c| c| c| c| c|} 
			\hline
			Material&SG& a(A$ {^0})$ & b(A$ {^0})$ & c(A$ {^0})$& PF& $R_f$\\ [0.5ex] 
			\hline
			(LCu)  & I4/mmm &3.78(9) & 3.78(9) &13.20(5)& NA&1.35\\ 
			\hline
			(LCo850) & R-3c &5.44(5) & 5.44(5) &13.22(5)& NA&1.44 \\
			\hline
			(LCo900) & R-3c &5.44(6) & 5.44(6) &13.21(7)& NA&1.44 \\
			\hline
			(LCo950) & R-3c &5.44(3) & 5.44(3) &13.20(5)& NA&1.44 \\
			\hline
			A1(LCu)&I4/mmm & 3.78(9)&3.78(9)&13.20(3)&76&1.47 \\
			\hline
			A1(LCo)&R-3c&5.44(2)&5.44(2)&13.20(9) &24&1.47 \\
			\hline
			A2/( LCu)&I4/mmm & 3.78(3) & 3.78(3) & 13.20(9) &76 & 1.44 \\
			\hline
			A2/( LCo)&R-3c & 5.44(6)&5.44(6)&13.21(5)&24&1.44 \\
			\hline
			A3/( LCu)& I4/mmm & 3.78(2) & 3.78(2) & 13.20(3)&76&1.8 \\
			\hline
			A3/( LCo)& R-3c & 5.44(1)&5.44(1)&13.21(7)&24&1.8 \\
			\hline
			A4/( LCu)& I4/mmm & 3.78(4) & 3.78(4) & 13.20(8)&86&1.6 \\
			\hline
			A4/( LCo)& R-3c&5.44(9)&5.44(9)&13.21(9)&14&1.6 \\
			\hline
			A5/( LCu)& I4/mmm&3.78(6)&3.78(6)&13.20(9)&95.9&1.87 \\
			\hline
			A5/( LCo)& R-3c&5.44(8)&5.44(8)&13.22(1)&4.1&1.87 \\
			\hline
		\end{tabular}
	\end{table}
		
	\begin{figure}[h]
		\centering
		\includegraphics[height= 4 cm, width=9.7 cm]{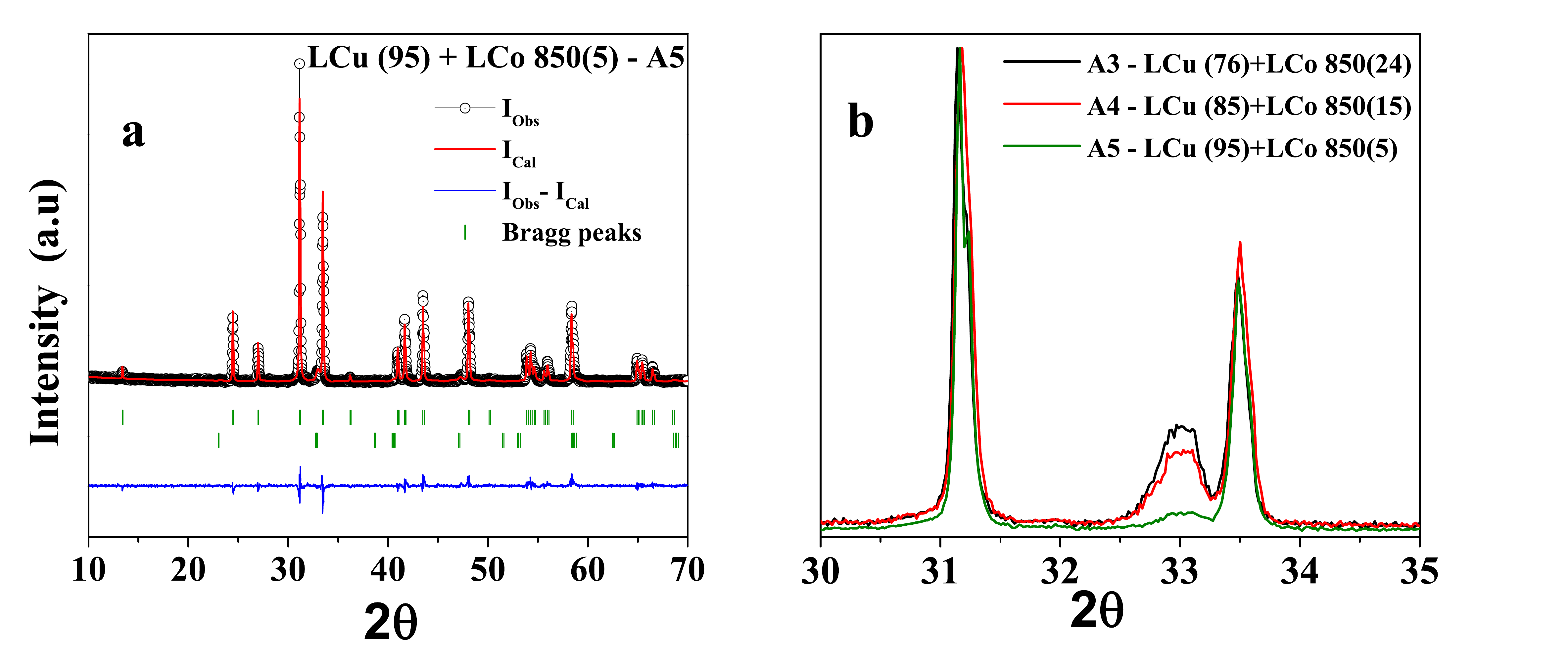}
		\vspace*{-8mm}
		\caption{(Colour Online) (a) Rietveld refinement of the composite A5. (b) XRD plot of all composites A3, A4, A5.}
		\label{fig:fig2}
	\end{figure}

	The lattice parameters of LCu obtained after the two phase Rietveld refinement of all composites (i.e., A1 to A5) do not show any significant change, which indicates the c/a ratio or the Orthorhombic distortion of LCu remain same for all composites. This indicates that there is no change in hole concentration of LCu \cite{r50}. The crystalline size of LCu and LCo is calculated using Williamson-Hall (W-H) analysis. The hole concentration of parent LCo is determined by Iodometric titration method. The chemical formula obtained from these results are as follows; like for LCo950, LCo900 and LCo850 it is $ La_{0.6} Sr_{0.4} Co O_{2.97}$, $ La_{0.6} Sr_{0.4} Co O_{2.99}$ and $ La_{0.6} Sr_{0.4} Co O_{3.01}$ respectively. From this chemical formula the concentration ratio of Co$^{+3}$ and Co$^{+4}$ ions can be calculated very easily and this ratio is further used to calculate the theoretical values of effective magnetic moment ($\mu_{eff}$) and saturation moment of the corresponding LCo. Experimentally the value of $\mu_{eff}$ has been calculated from the Curie-Weiss fitting in the paramagnetic region of LCo and the obtained value is compared with the theoretical effective magnetic moment value of LCo($(\mu_{eff})_{expt}$=4.33$\pm$0.01). As LCu is a Pauli paramagnet and the moment value above superconducting onset temperature is found around 10$^{-6}$ emu (much smaller than the Curie-Weiss moment value of LCo), and also found constant of temperature up-to room temperature. Therefore, the effective magnetic moment value of parent LCo and the corresponding composites has to be same provided there is no change of hole concentration or chemical reaction. We have used this concept to determine the spin state (i.e. hole concentration) of LCo present in the composite \cite{c4r24}. 
		
\subsection{Transmission electron microscopy (TEM)}
\vspace*{-2mm}
Transmission electron microscopy (TEM) image of composite A3 is shown in \texttt{FIG.\,3(a)}. The high resolution transmission electron microscopy (HRTEM) image is shown in \texttt{FIG.\,3(b)}, \texttt{FIG.\,3(c)} and \texttt{FIG.\,3(d)}. The high resolution images are depicting that the crystallites of LCo are connected and combined with the crystallite of LCu, respectively. The fringes of the marked lattice spacing of 0.5 nm, corresponds
to LCo. \texttt{FIG.\,3(d)} indicates very sharp interface, which clearly depicts there is no intermixing between LCu and LCo. We have performed similar TEM and HRTEM measurements at various places of A3 and also performed High angle annular dark field imaging (i.e., elemental scan), ( see Section.\,A  in the Supplemental Material below for more elaborate explanation).

\begin{figure}[h]
	\centering
	\includegraphics[width=8.5 cm, height= 7 cm]{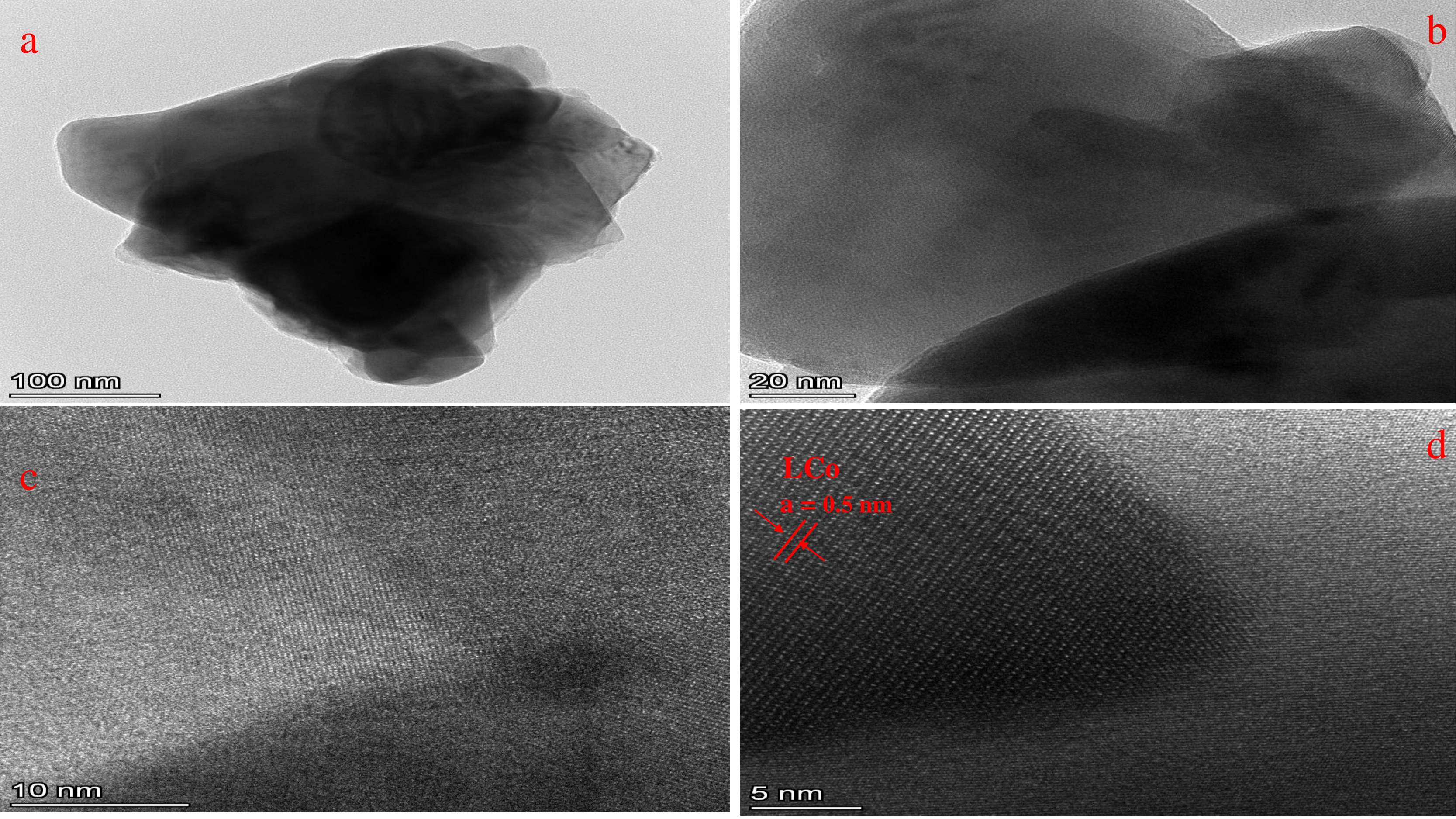}
	\vspace*{-2mm}
	\caption{(Color Online) HRTEM image of composite A3 is shown at various resolution. }
	\label{fig:fig3}
\end{figure}

\maketitle\section{EXPERIMENTAL DETAILS}
	\vspace*{-2mm}
	The Low field linear and nonlinear magnetic ac-susceptibility measurements have been performed using a homemade ac-susceptometer, which can be operated down to 4\,K from 300\,K and the measurements can be done in both cooling and heating cycle with a temperature accuracy better than 1\,mK. The estimated sensitivity of the setup is $\sim 10^{-7}$\,emu \cite{r39}. The higher dc-field($>$ 200\,Oe) superimposed ac-susceptibility measurements are performed in MPMS-XL (M/S, Quantum Design).
		\vspace*{-5mm}
	\subsection{Ac-susceptibility measurement}
	\vspace*{-3mm}

	The magnetic property of the composite has been studied by extensive use of ac-susceptibility measurement which is capable to probes the spin dynamic at very low field. The magnetization (m) can be expanded with respect to the applied ac-field $h_{ac}(h) $ as
	\[ m = m_0+\chi_1 h+ \chi_2 h^2+\chi_3 h^3+\chi_4 h^4....(1)\]
	$\chi_1(\approx\delta m/\delta h$) is linear susceptibility and  $\chi_2$, $\chi_3$, $\chi_4$...  are nonlinear susceptibilities. These nonlinear susceptibilities contain many fruitful information but magnitude of these are much smaller (couple of order) than the linear susceptibility, therefore they are difficult to measure from normal dc magnetization measurement, but these nonlinear susceptibilities can be easily measured from high sensitive ac-susceptibility measurement  \cite{r39a}. If the magnetization has an inversion symmetry with respect to the applied ac-field ($h_{ac} $) then all the even order susceptibilities, like $\chi_2(\approx\delta^2 m/\delta^2 h$), $\chi_4(\approx\delta^4 m/\delta^4 h$) are zero without any externally applied dc-field ( i.e. at h$_{dc}$\,=\,0\,Oe) like paramagnetic and antiferromagnetic materials, but when the  inversion symmetry breaks w.r.t the sign of $h_{ac} $ then  $\chi_2$, $\chi_4$.., all the even order susceptibility shows finite value (at h$_{dc}$\,=\,0\,Oe), like in case of ferromagnet, or ferrimagnetic state\cite{r39a,x4,x6,x7}.
	
	Third order susceptibility ($\chi_{3}$) is very useful tool to discriminate between various metastable states (like spin glass (SG), superparamagnet (SPM) etc.) ably \cite{c4r41,c4x1,c4x2,c4r47,c4r48}. For example, if there are inter-particle interaction then the frequency dependent nature of $\chi_{1}^R$(T) for a SPM system will give similar result as that of a SG system. Hence, in this type of situation it becomes difficult to discriminate between SG and SPM by looking only at the frequency dependent $\chi_{1}^R$(T)  result \cite{c4r47}. Whereas across spin glass transition a negative cusp in $\chi_{3}$ is observed which is correlated with the divergence of Edward-Anderson order parameter, but in case of SPM this type of divergence is absent \cite{c4r41,c4x1,c4x2}. $\chi_{3}$ is also found as very effective tool in determining the universality class of  the ferromagnet i.e. to study the nature of magnetic ground state of the corresponding ferromagnet \cite{x4,c4r48a}. Therefore, the behavior of  $\chi_{3}$ around ferromagnetic to paramagnetic transition (T$_C$) is very important in deciding the nature of magnetic interaction or the nature of magnetic ground state of a ferromagnet. 

\maketitle\section{Results and discussions}	
\vspace*{-3mm}
	The combined temperature-dependent plot of mole normalized $\chi_1^R$(T) (the real part of first order susceptibility) of A1, A2 and A3 as well as  their respective ferromagnetic constituent is shown in \texttt{FIG.\,4(a)}. $\chi_1^R$(T) of the composite A1 is indicated by down red triangle and $\chi_1^R$(T) plot of the corresponding ferromagnet LCo950 is indicated by Red line. Similarly black and blue color (symbols and line) describes the same for composite A2 and A3 respectively. All measurements are performed in heating cycle. Normalization of $\chi_{1}^R$(T) is performed with respect to the mole fraction of LCo as obtained from the Rietveld refinement of the XRD patterns \cite{c4r24}. The Zoomed view of $\chi_1^R$(T) around FM T$_C$ is shown in the upper inset of \texttt{FIG.\,4(a)}, it is observed that around T$_C$ and above it  the $\chi_1^R$(T) value of the composite matches with the $\chi_1^R$(T) value of the corresponding ferromagnet present in that. These observation indicates  that the Curie temperature ($\theta_C$) and effective magnetic moment ($\mu_{eff}$) of the composites and the corresponding parent LCo are same ($\theta_C$ and $\mu_{eff}$ have been calculated from the Curie-Weiss fitting of the paramagnetic region of $\chi_{1}^R(T)$ graph \cite{c4r24}). It suggests the magnetic structure (or oxygen stoichiometry) of LCo in the composites remain same with respect to their parent LCo. Similar conclusion has been drawn from $\chi_{3}$ as shown later (\texttt{FIG.\,5}).  Hence the anomaly around the superconducting onset temperature, $T_{S(onset)}$, shown in the lower inset of \texttt{FIG.\,4(a)} indicates the modification of  magnetic state of LCu due to close proximity of LCo. The anomaly is more prominent in case of A3 than A2 and A1, because A3 consists of smaller ferromagnetic particle LCo850, hence it has larger surface to volume ratio than LCo900 and LCo950, due to this reason the effective interface between LCu and LCo is larger in case of A3 than A1 and A2. Therefore the above observation depicts the anomaly around T$_{S(onset)}$ is an interface effect. 	
	\begin{figure*}[t]
		\centering
		\includegraphics[width=16 cm, height= 7 cm]{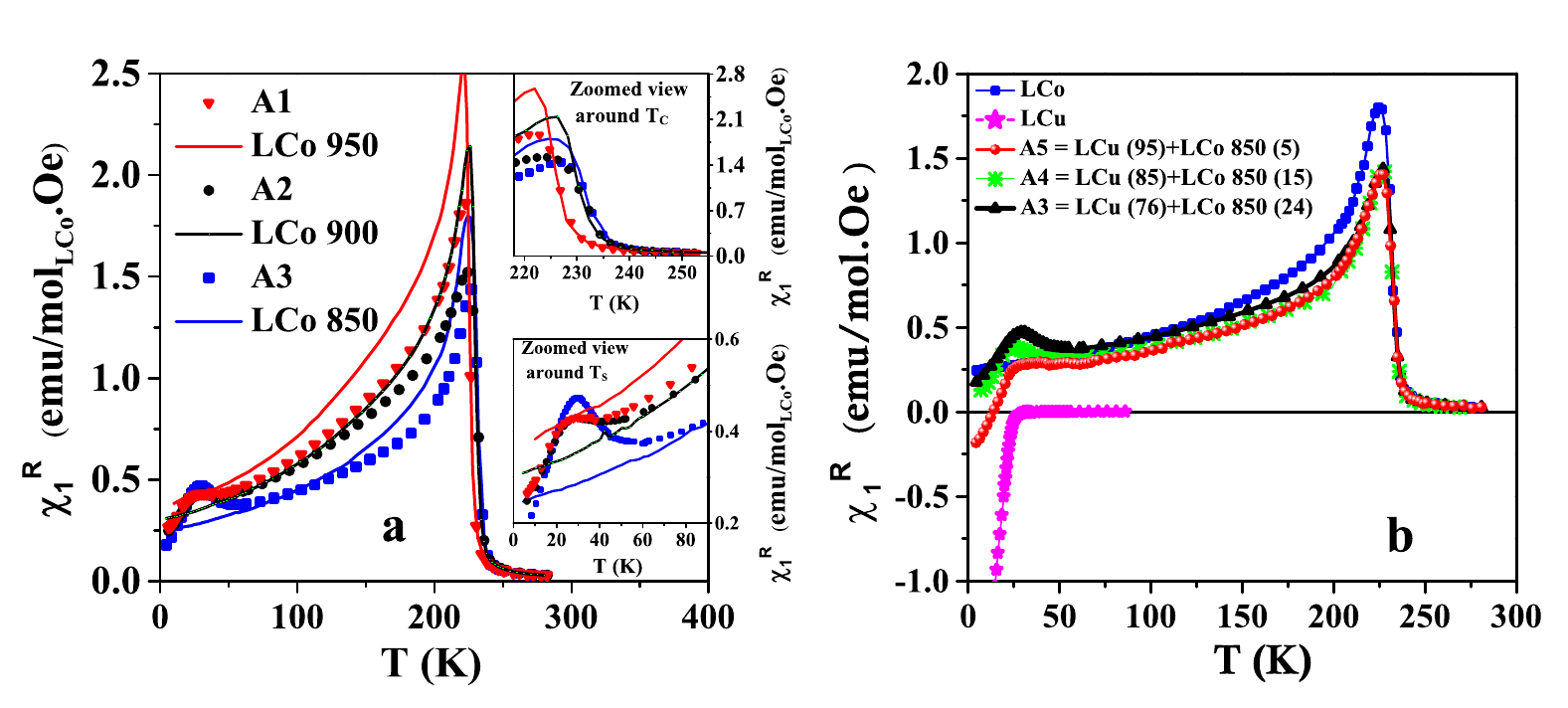}
		\vspace*{-2mm}
		\caption{(Color Online) (a)The combined temperature dependent plot of $\chi_{1}^R(T)$ for the composite A1 (red triangle) and LCo950 (Red line), composite A2 (black circle) and LCo900 (black line),  composite A3 ( blue rectangular box) and LCo850(blue line). Upper inset shows the zoomed view around ferromagnetic to paramagnetic transition temperature(T$_C$). Lower inset shows the zoomed view around superconducting onset temperature (T$_{S(onset)}$). (b) The combined temperature dependent plot of A3, A4, A5, LCu and LCo depicted. All the measurements are done in an ac-field of amplitude 3\,Oe and frequency 231.1\,Hz.}
		\label{fig:fig4}
	\end{figure*}

	To cross check the statement related to the interface effect, the ratio of LCu and LCo is changed in composites A4 and A5 (as mentioned before) to vary the number of LCo grains sprinkled across LCu domain, which also changes the effective interface between LCu and LCo. {FIG.\,4(b)} shows the comparative normalized $\chi_{1}^R$(T)  plot of composites A3, A4 and A5, along with them the normalized $\chi_1^R$(T) graph of parent LCo and LCu is also shown for comparison purpose. The onset temperature of superconductivity ($T_{S(onset)}$) of LCu  is observed around 32\,K and above $T_{S(onset)}$, $\chi_1^R(T)$ shows Pauli paramagnetic behavior. The paramagnetic to ferromagnetic transition temperature (T$_C$) of LCo850 is observed around 233\,K. The anomalous hump in $\chi_{1}^R$(T) around the superconducting transition gets suppressed with the reduction of LCo850 concentration in LCu matrix (shown in {FIG.\,4(b)}), like for A4 the amplitude of the peak value suppresses and become almost flat for A5.  From Curie-Weiss fitting in the paramagnetic region the value of $\mu_{eff}$ and $\theta$ is obtained around 4.28\,$\mu_b$ and 233\,K respectively. The Hopkinson peak amplitude (i.e the peak in $\chi_{1}^R$(T) graph just below FM T$_C$) for all the composites remain same, which indicates the magnetic domain size of LCo is same in all composites. Thus, the observed anomaly suppresses in similar manner as that of {FIG.\,4(a)}, which is an indication that the anomaly in $\chi_{1}^R$(T) measurement is an interface driven phenomena.
	
	\begin{figure}[b]
		\centering
		\includegraphics[width=9.5cm,height=8cm]{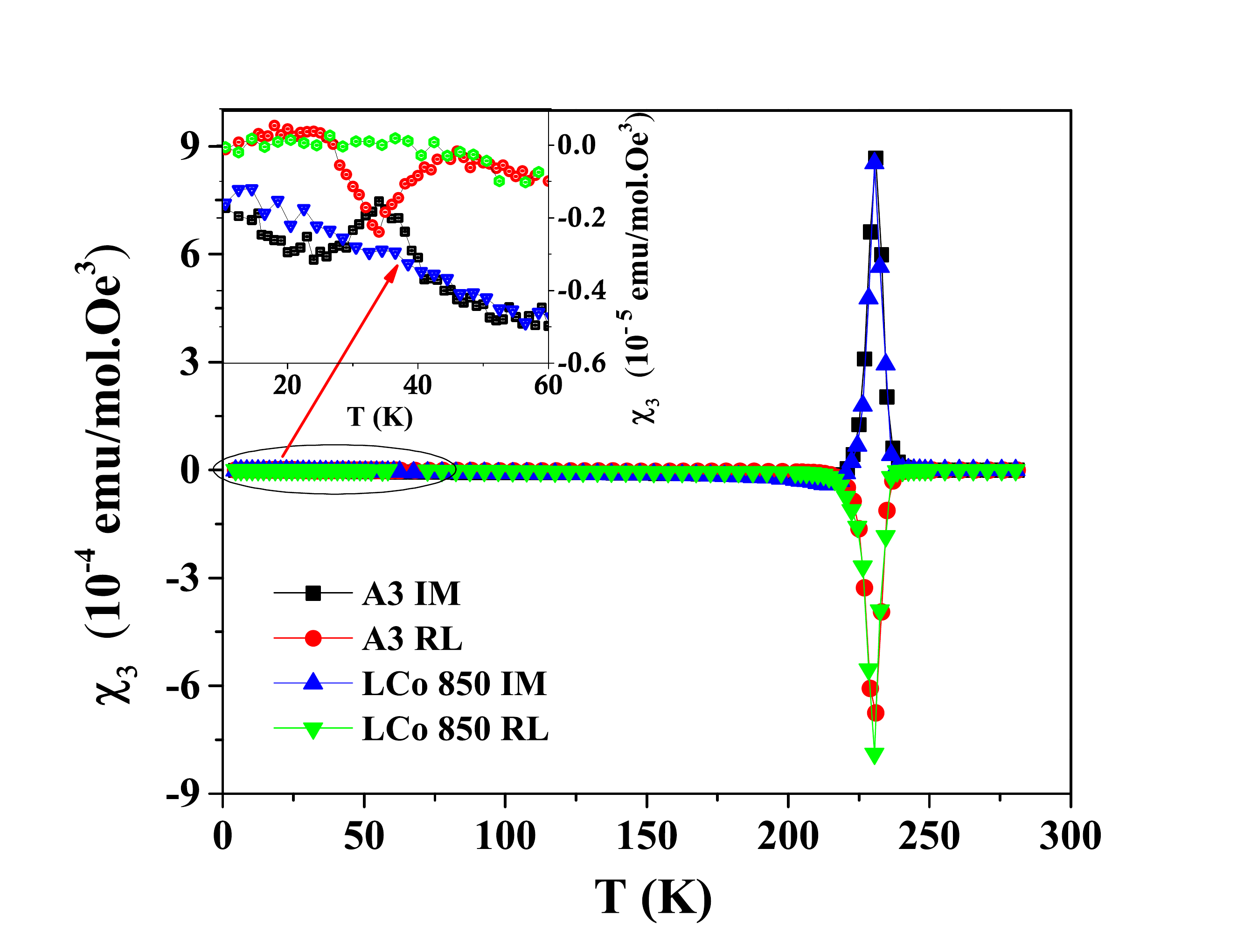}
		\vspace*{-2mm}
		\caption{\textit{\small{(Color Online) The temperature dependent mole normalized Real (RL) and Imaginary (IM) parts of  $\chi_3$ of Composites A3 and LCo850. Inset shows the zoomed view of the anomalous region around superconducting onset (T$_{S(onset)}$). The measurements are performed at ac-field($h_{ac}$)= 3 Oe and frequency(f)= 231.1\,Hz.}}}
		\label{fig:fig5}
	\end{figure}
	
	The mole normalized $\chi_{3}$ plot of the composite A3 and the corresponding ferromagnet is shown in FIG.\,5 (the number of mole of LCo present in the composite is used to normalize the $\chi_{3}$ graph of the composite). It shows that the nature of $\chi_{3}$ (both real and imaginary parts) remain same for the composite and the ferromagnet except around the superconducting onset temperature (T$_{S(onset)}$), which gives direct indication that the magnetic state of the ferromagnet present in the composite remain same as that of the parent ferromagnet.

	The interface effect is further verified by measuring the $\chi_{1}^R(T)$  behavior of a composite pallet (A2) and the powder specimen of same composite; the comparative plot is shown in {FIG.\,6}. The inter-granular connection reduces after making powder of the composite pallet and hence the effective interface between LCu and LCo reduces. The difference between the  $\chi_{1}^R(T)$ plot of powder and pallet around $T_{S(onset)}$ can be clearly distinguished from the inset of {FIG.\,6}. For powder the hump is suppressed as a result of reduction of effective interface but there is no change in the critical temperature of the ferromagnet. So all these measurements are unambiguously show that when the effective interface reduces then the anomaly around $T_{S(onset)}$ also decreases, which clearly proves the phenomena is an interface effect. 

	\begin{figure}[t]
		\centering
		\includegraphics[width=8cm]{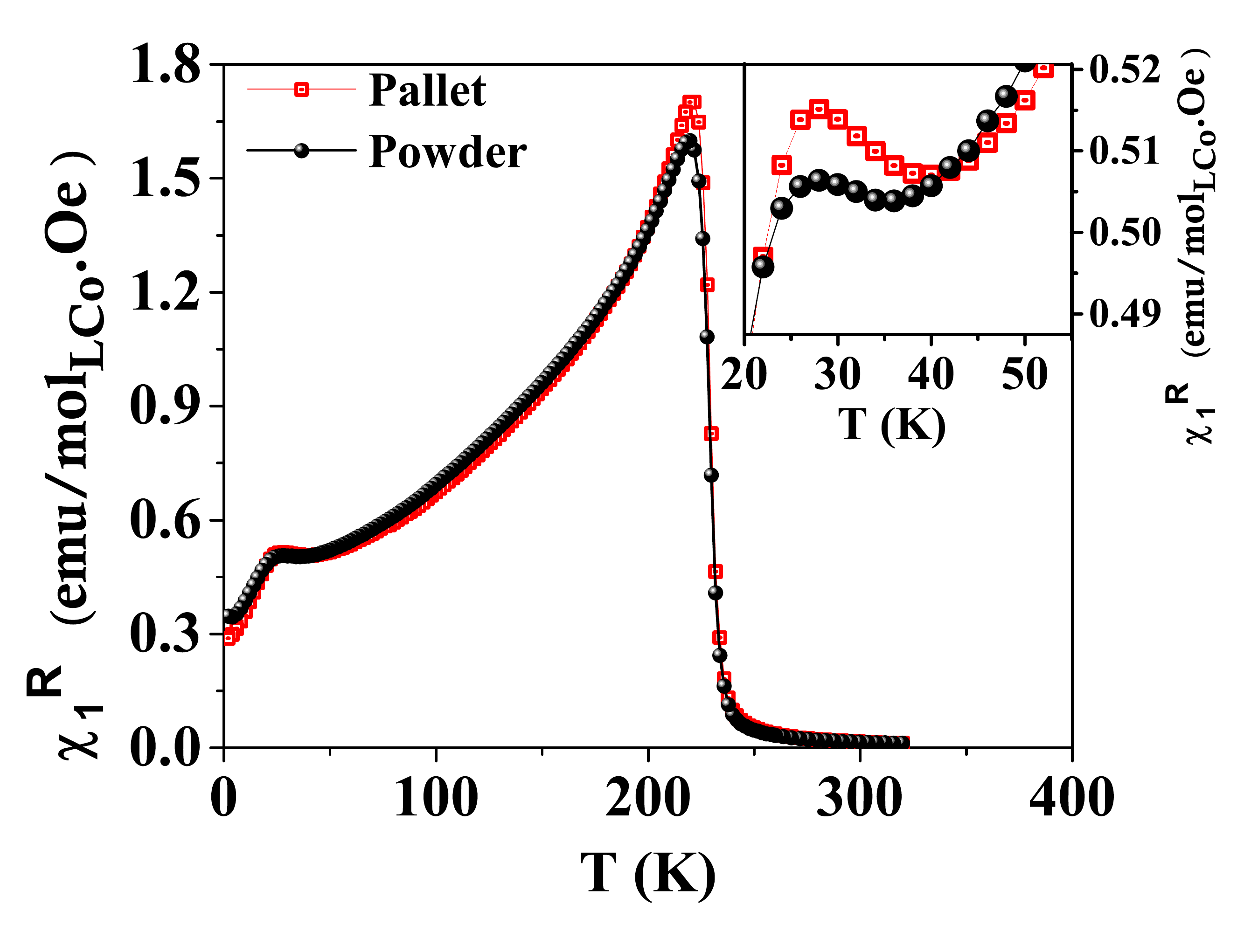}
		\vspace*{-5mm}
		\caption{(Color Online) The plot of  $\chi_{1}^R(T)$ against temperature for A2 composite in two different conditions. The red square block shows the result for pallet and the black circle for the powder samples. Inset shows the zoomed view of the anomalous region. The measurements are performed at ac-field($h_{ac}$)= 3 Oe and frequency(f)= 231.1\,Hz.}
		\label{fig:fig6}
	\end{figure}

	The gradual decrease of the diamagnetic fraction below $T_{S (onset)}$ while increasing the effective interface is evident from {FIG.\,4(b)}. The crystallite size and the lattice parameter of LCu (and LCo) remain almost same for all composites (evident from the XRD graph). The magnetic ac-susceptibility measurement depicts the demagnetization factor and spin state of LCo also remain same for all the composites (evident from the equal values of the Hopkinson peak height at 226\,K and paramagnetic moment), so the decrement of diamagnetic phase fraction is not related to the degradation of hole concentration or structural distortion of either LCu or LCo. This indicates a possibility that there has developed some kind of magnetic interaction between LCu and LCo across the interface, which causes the observed changes in the magnetic state of LCu (like, the decrease in diamagnetic susceptibility and excess susceptibility across $T_{S (onset)}$). Otherwise, if it is due to the change of magnetic state of LCo then the change could have been observed around T$_C$ with increasing or decreasing concentration of LCo (in FIG.\,4(b)). Along with that, above T$_{S(onset)}$ the magnetization strength of LCu (i.e. Pauli paramagnetic) is much smaller than LCo (ferromagnetic), which can not affect the magnetic state of LCo but the vice versa is highly probable. Hence, it can be concluded that the magnetic state of LCu is getting affected because of its proximity to LCo, without any change in the chemical composition of the constituents in the composites.

	To reveal the nature of perturbed magnetic state of LCu and also the type of magnetic interaction across the interface, the higher order (i.e. second order-$\chi_2$ and third order-$\chi_3$) ac-susceptibility  measurements are  performed because of their uniqueness to distinguish between various magnetic state (as previously discussed). Around T$_{S(onset)}$,  $\chi_{3}^R$ shows a dip like feature (shown in the inset of FIG.\,5) and around the similar temperature range  $\chi_{1}^R$ shows hump like behavior (as shown in FIG.\,4). This  behavior is evocative of $\chi_{3}^R$ as observed in spin glass (SG) \cite{c4x1,c4r41,c4r42} and superparamagnetic (SPM) systems \cite{c4x2,c4r44,c4r45}.	But unlike SG in this case $\chi_{3}^R$ doesn't show any critical behavior with respect to applied magnetic field, frequency and temperature. Moreover, the temperature dependent behavior of  $\chi_{1}$ and $\chi_{3}$ follow Wohlfarth model \cite{c4r44,c4r45,c4r47,c4x2}. The ac-field dependent behavior of $\arrowvert\chi_3^R\arrowvert$ is shown in the right hand side upper inset of FIG.\,7(a) and the value of $\arrowvert\chi_3^R\arrowvert$ approaches towards saturation with decreasing the amplitude of ac-field, which is a clear indication of the blocking phenomena of magnetic clusters. According to Wohlfarth SPM model \cite{c4r44} the magnetization(M) of a non interacting single domain SPM particle is represented as 
	\begin{equation}
		M = n\bar{\mu}L(\frac{\bar{\mu}}{k_BT})
	\end{equation}
	where n is the number of particles per unit volume, $\bar{\mu}$ is the average magnetic moment of a single magnetic entity or particle, $k_B$ is the Boltzmann’s constant, and L(x) is the Langevin function. So, from \texttt{Eqn.\,1} the linear susceptibility can be expressed as
	\begin{equation}
		\chi_{1}^R=\frac{\bar{\mu}}{3k_BT}
	\end{equation}
	and the third order susceptibility is represented as  
	\begin{equation}
		\chi_{3}^R=-(\frac{n\bar{\mu}}{45})(\frac{\bar{\mu}}{k_BT})^3
	\end{equation}
	
	\begin{figure*}[t]
		\centering
		\includegraphics[width=17cm,height=7cm]{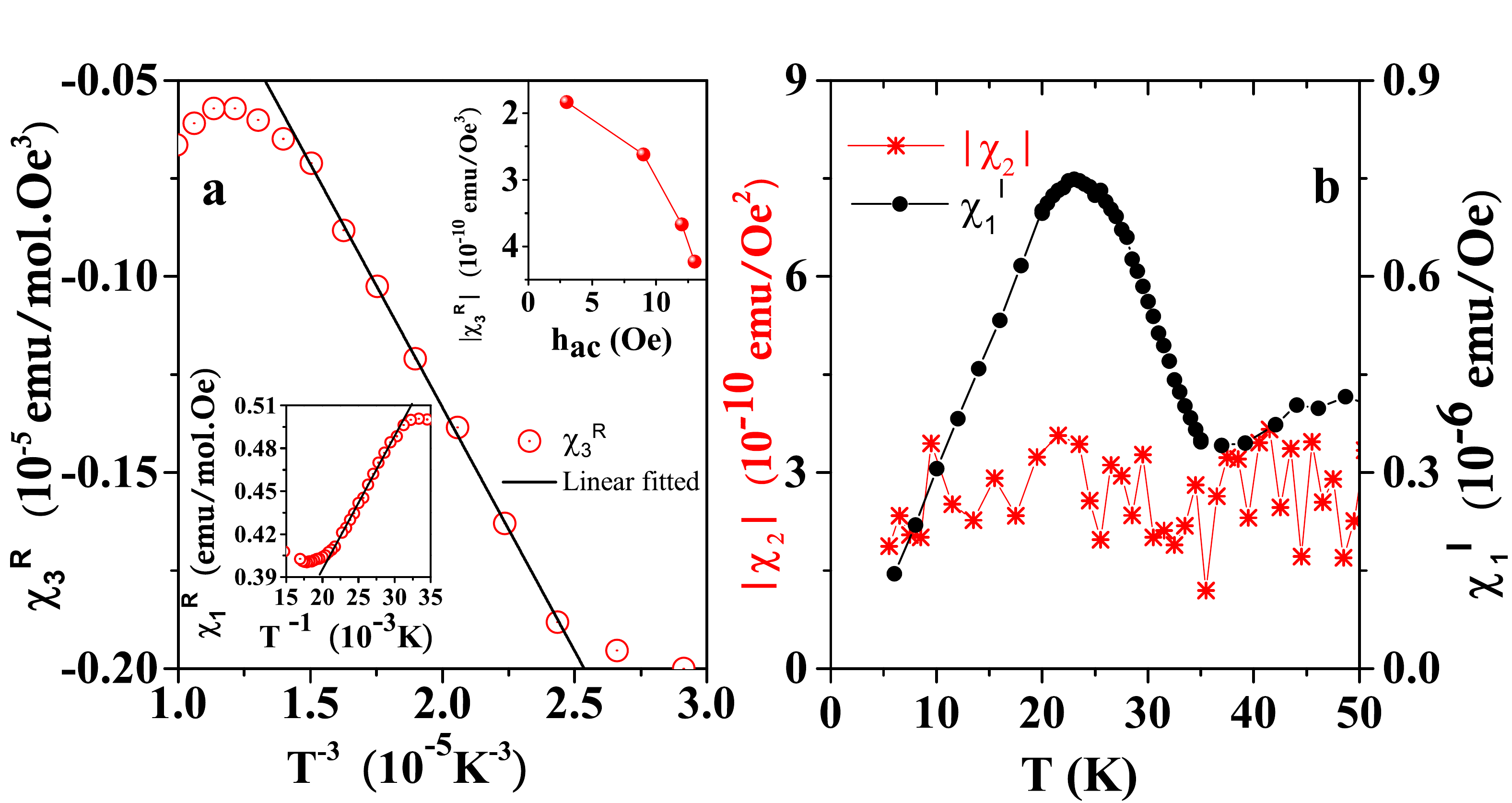}
		\vspace*{-3mm}
		\caption{\textit{\small{(Colour Online) (a) $\arrowvert\chi_3^R\arrowvert$ is plotted against $T^{-3}$ and the black line shows the linear fitting. Inset (Top Right) $\arrowvert\chi_3^R\arrowvert$ is plotted  against amplitude of applied ac-magnetic field at temperature 33\,K for the sample A3. Inset (Down Left) $\chi_1^R$ is plotted against $T^{-1}$ above 32K black line shows the linear fitting to the data. (b) Imaginary part of $\chi_1^I$(Right hand side Y axis) and  $\arrowvert\chi_2\arrowvert$ (left hand side Y axis) are plotted  against temperature of A3 composite. The measurements are performed at $h_{ac}$\,=\,12\,Oe at an exciting frequency of (f) = 231.1 Hz. Only in the Top-Right inset of (a) the the exciting ac-field is varied at same frequency.}}}
		\label{fig:fig7}
	\end{figure*}

	Therefore, \texttt{Eqn.\,2} and \texttt{Eqn.\,3} demonstrates that in the SPM region $\chi_{1}^R$ is positive and it follows $T^{-1}$ behavior, whereas $\chi_{3}^R$ is negative and it follows $T^{-3}$ behavior \cite{c4r44}. FIG.\,7(a) shows the $T^{-3}$ temperature dependence for experimentally obtained $\chi_{3}^R$ and lower inset of FIG.\,7(a) shows the $T^{-1}$ dependency of $\chi_{1}^R$ above 32\,K. The value of  $\bar{\mu}$ ($\sim 10^3$\ $\mu_b$, $\mu_b$ is Bohr magneton) is calculated from the slope ratio obtained from the straight line fitting of $\chi_{3}^R$ and $\chi_{1}^R$. In case of a conventional SPM cluster (smaller particles of ferromagnet) the values of  $\bar{\mu}$ are quite large (as it contains large number of ferromagnetically align spins inside its volume, 	for a pure ferromagnetic clusters it is of the order of 10$^4$-10$^6$$\mu_{b}$ \cite{c4r47}) compared to the number we have obtained from fitting. It indicates that the SPM clusters are not ferromagnetic in nature. Moreover, there is no anomaly in the temperature dependent $\arrowvert\chi_2\arrowvert$ measurement is observed through out the temperature interval as shown in FIG.\,7(b) (scale: Left hand Y axis scale), indicating zero value of internal field (i.e. non-ferromagnetic type clusters). FIG.\,7(b) also shows the temperature dependent plot of $\chi_{1}^I$ (scale: right hand Y axis scale), below 33\,K a hump like features is observed in $\chi_{1}^I$ graph and no further anomaly in $\chi_{1}^I$ is observed above 33\,K. $\chi_{1}^I$ corresponds the imaginary part of first order ac-susceptibility and signifies area of minor magnetic hysteresis loop \cite{c4r27}, which does not show any anomaly both around and above blocking temperature T$_B$. Therefore all these observations are ruling out the possibility of forming ferromagnetic type of clusters around T$_{S(onset)}$. In ac-susceptibility measurement at zero dc bias field  the AFM type ordered systems (or clusters) show zero value of internal field (i.e.$\arrowvert\chi_2\arrowvert$$=$0) and null value of minor hysteresis loop area (i.e. $\chi_{1}^I$=0) \cite{c4r48}, which is similar to our observations. Therefore, the fundamental and higher order susceptibility measurement confirms the excess susceptibility or hump in $\chi_{1}^R$ around T$_{S(onset)}$ appears due to the blocking of the SPM type AFM clusters.
		
		\begin{figure*}[t]
		\centering
		\includegraphics[width=17cm,height=7cm]{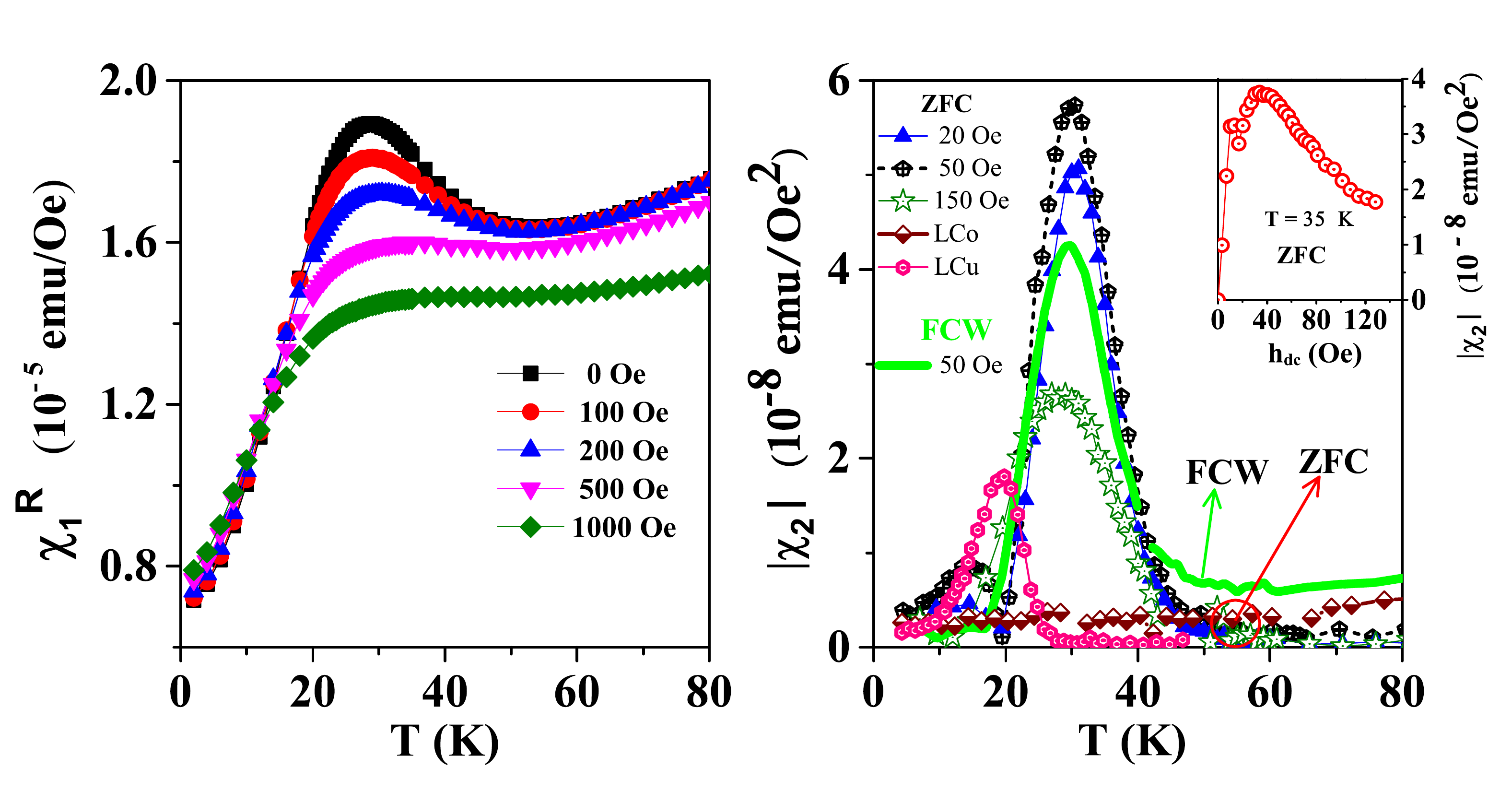}
		\vspace*{-6mm}
		\caption{\textit{\small{(Color online) (a) $\chi_{1}^{R}$ data of A3 is plotted against temperature in five superimposed dc-field 0, 100, 200, 500 and 1000\,Oe (b) $\arrowvert\chi_2\arrowvert$ graph of A3 is shown in three superimposed dc-fields 20, 50 and 150\,Oe, The measurements are performed in ZFC and FCW (Green bold line) along with that the plots for LCu and LCo are also shown at 50\,Oe superimposed dc-field in ZFC mode. Inset shows the isotherm of $\arrowvert\chi_2\arrowvert(H)$ of A3 at T=35\,K. For all measurement the amplitude and frequency of the ac-field are $h_{ac}$=12\,Oe and f=231.1\,Hz.}}}
		\label{fig:fig8}
	\end{figure*}

	SPM clusters are short range ordered magnetic clusters and its anisotropy energy is comparable to the thermal energy. Due to this reason the clusters show dynamical behavior above blocking temperature but the microscopic spin structure or the interaction between the spins remain same to that of long range order magnetic structure or state \cite{c4x2}. These short range ordered AFM clusters can appear from the covalent bonding between $Cu^{+2}$ ions and $Co^{+3}$ or $Co^{+4}$ ions across the interface, as it is observed in YBCO/LCMO heterostructure \cite{c4r10,c4r11}. But in our case the large suppression of the diamagnetic fraction (shown in FIG.\,4(b)) along with the significant change of $\chi_{1}^R$(T) around $T_{S(onset)}$ ($\sim10^{-5}$\,emu, shown in FIG.\,4) indicates the perturbation is not only confine to the interface but it is propagated well inside the bulk of LCu i.e. the bulk magnetic property of LCu is modulated due to close proximity of LCo. Recently, the emergence of bulk CDW phase is evident in cuprate/ferromagnet heterostructure \cite{c4r13,c4r14} and the appearance of similar CDW phase, AFM phase, AFM glass phase is also evident while the superconductivity of a cuprate is destroyed by applying very high magnetic field or by applying very high strain on it \cite{c4r50,c4r51,c4r52,c4r53,c4r54,c4r55,c4r56}. Here we have observed formation of the short range ordered SPM type AFM clusters along with decrement of diamagnetic fraction which are unrelated to change in crystal structure or change of charge carrier concentration. Hence, the most probable reason is, the AFM fluctuation of LCu is suppressed by the exchange magnetic field of LCo. The exchange bias field can propagate through the covalent bonding between $Cu^{+2}$ ions and $Co^{+3}$ or $Co^{+4}$ ions across the interface or directly inside LCu, and resulted in formation of short range order SPM type AFM clusters both across the interface and also within the bulk of LCu.
			
	The presence of AFM type spin fluctuation in optimally doped LCu has already been found from various measurements like specific heat, neutron diffraction, X-ray scattering \cite{c4t2,c4r1,c4r2,c4m1,c4r50,c4r51,c4r52,c4r58,c4r59,c4r60} etc.  In this section we are going to emphasize on the phenomena related to exchange magnetic field stabilized  AFM ordering in cuprate superconductor. For that a small dc bias field is superimposed with the ac-field during the susceptibility measurements. FIG.\,8(a) shows the dc-field superimposed $\chi_{1}^R$(T) plot of composite A3, the peak value of $\chi_{1}^R$(T) suppress and the peak temperature (corresponding to the blocking temperature T$_B$) shifts towards higher temperature ($\Delta$(T$_B$) $\sim$ 6\,K) with increasing the amplitude of dc bias field. The applied maximum dc bias filed (h$_{max}$) is very small in magnitude, hence the Zeeman energy does not have sufficient strength to disturb the already formed AFM clusters in LCu, but the magnetic response of LCo can be manipulated. At very lower value of dc bias field, the loosely coupled surface spins are responding and when the dc-field amplitude exceeds the value of demagnetization field of the ferromagnet ($\sim$50\,Oe), then spins at the bulk of LCo are responding, because above the demagnetization field Zeeman energy overcome the magneto crystalline anisotropy energy and hence the cobalt spins align along a resultant direction decided by the competition between these two interactions. As the cobalt spins are coupled with the copper spins by exchange force, therefore the effective exchange bias field on LCu created by LCo also changes due to the change of orientation of cobalt spins. Thus, the preformed AFM clusters feel this change and get biased indirectly by this small value of applied dc-field, and hence the decrement of relaxation peak height of $\chi_{1}^R(T)$ and the positive increment of T$_B$ with increasing the amplitude of dc bias field is observed.

		To further emphasize on this exchange bias model the dc-field superimposed second order susceptibility ($\chi_2$)  measurements are performed, as the application of dc bias field tunes the amplitude of effective exchange bias field seen by the copper spins of LCu. The copper spins are at different distances from the interface hence, each spins of LCu is going to feel different amplitude of exchange force created by the cobalt spins. So, the internal spin arrangement of the antiferromagnetic clusters in LCu is not going to remain same as the previous case of h$_{dc}$=0\,Oe. Therefore, these change of the internal field amplitude is going to reflect on the second order susceptibility ($\arrowvert\chi_2\arrowvert$). The dc-field superimposed temperature dependent $\arrowvert\chi_2\arrowvert$ measurement ($\arrowvert\chi_2\arrowvert$(T)) is shown in FIG.\,8(b). The anomaly in $\arrowvert\chi_2\arrowvert$(T) appears around 50\,K and persists up to 5\,K. These measurements are performed in ZFC (zero field cooled) mode, the sample is cooled in zero field and all measurements are performed in warming cycle in the presence of dc-field ( $h_{dc}$= 20, 50 and 150\,Oe respectively). One of the graph in FIG.\,8(b) (Green bold line) corresponds to the FCW (field cool warming) measurements taken at $h_{dc}$=\,50\,Oe. In this protocol dc-field is applied above $T_C$ of A3, then cooled to 5\,K and measurements are performed during warming cycle in the presence of same dc-field (50 Oe). In ZFC measurement initially the peak value of $\arrowvert\chi_2\arrowvert(T)$ (FIG.\,8(b)) increases up to dc-field of amplitude 50\,Oe and then it decreases with further increasing the amplitude of the dc bias field. Inset shows the isotherm of  $\arrowvert\chi_2\arrowvert$ (i.e.$\arrowvert\chi_2\arrowvert(H)$) at 35\,K, revealing the same details as obtained from $\arrowvert\chi_2\arrowvert(T)$, viz  $\arrowvert\chi_2\arrowvert$ decreases after a particular field (50\,Oe). Similar measurements are performed for the parent ingredients, but no anomaly in $\arrowvert\chi_2\arrowvert$  is observed around these temperature(i.e. 20\,K-50\,K). The $\arrowvert\chi_2\arrowvert$(T) graph of LCu depicts the appearance of $\arrowvert\chi_2\arrowvert$ below about 25\,K (In case of cuprate superconductor magnetic field dependent critical current density is the source of $\arrowvert\chi_2\arrowvert$  \cite{c4r28,c4r29,c4r30}) and in case of LCo $\arrowvert\chi_2\arrowvert$ appears around $T_C$ ($\sim$233\,K, because of the appearance of internal symmetry breaking field but much below $T_C$, $\arrowvert\chi_2\arrowvert$ shows almost zero value because of the domination of demagnetization effect), so neither LCo nor LCu alone is responsible for the anomaly observed in $\arrowvert\chi_2\arrowvert$ around 35\,K for the composites.

	This anomalous variation of $\arrowvert\chi_2\arrowvert$ with dc-field can be very nicely explained by using the exchange coupling model. The mathematical form to describe exchange coupling between two magnetic material across the interface is as follows \cite{c4r61,c4r62,c4r63,c4r64}
	\begin{equation}
		H_{ex} =-j\frac{S_{AFM}\times S_{FM}}{\mu\times t_{FM} \times M_{FM}}
	\end{equation}\
	$H_{ex}$ is the exchange bias field, J is the exchange integral across the FM/AFM interface per unit area, $S_{AFM}$ and $S_{FM}$ are the interface (or surface) magnetization amplitude of the antiferromagnet and the ferromagnet, respectively, $t_{FM}$ is the thickness of the ferromagnetic domain and $M_{FM}$ is the magnetization of the FM layer at a magnetic filed `h'. At lower value of dc bias field ($<$50\,Oe), the spins across the interface are only going to respond (i.e. $S_{FM}$, as they are loosely coupled to the bulk of LCo), whereas the bulk magnetization are not going to respond at this lower value of dc-field i.e. $M_{FM}$ and $t_{FM}$ remain unchanged (because the applied field amplitude is still lower than the values of demagnetization field). Therefore, according to \texttt{Eqn.4} at lower dc-field amplitude the effective $H_{ex}$ (Local field) on copper spins is larger in amplitude, which causes comparatively sizable misalignment between the AFM coupled copper spin, as they are at different distance from the interface. Due to this reason the increment of $\arrowvert\chi_2\arrowvert$ is observed up-to a certain value of dc magnetic field (i.e. 50\,Oe). On further increasing the dc-field amplitude the bulk spins start to respond, because it exceeds the amplitude of demagnetization field of LCo (i.e., $M_{FM}$ and $t_{FM}$ increases simultaneously), as a result $M_{FM}$ and $t_{FM}$ start to dominant over $S_{FM}$, and hence, $H_{ex}$ reduces in accordance with \texttt{Eqn.\,4}. As a result the misalignment between the copper spins reduces and decrement of $\arrowvert\chi_2\arrowvert$ is observed at higher value of dc bias field. Similarly, the difference of $\arrowvert\chi_2\arrowvert$(T) values between ZFC and FCW protocol (shown in FIG.\,8(b)) can also be explained by using \texttt{Eqn.\,4}. As the ferromagnetic domain size ($t_{FM}$) and $M_{FM}$ of LCo in FCW condition is larger than the ZFC condition (below saturation field), evident from higher value of $\arrowvert\chi_2\arrowvert$ in case of FCW than ZFC above 50\,K, i.e. higher value of internal field, hence the resultant exchange bias field values on copper spins is larger in ZFC case than FCW case. As a result the observed values of $\arrowvert\chi_2\arrowvert$(T) below 50\,K  is lower in FCW case than it is in ZFC case. This observation further supports the proposed exchange bias model. Additionally, a measurement of the minor magnetic hysteresis loop, which is discussed in the Supplemental Material (See Section.\,B), also shows evidence of an exchange bias between the Copper and Cobalt spins.
	
	\begin{figure*}[t]
		\centering
		\includegraphics[width=17cm,height=7cm]{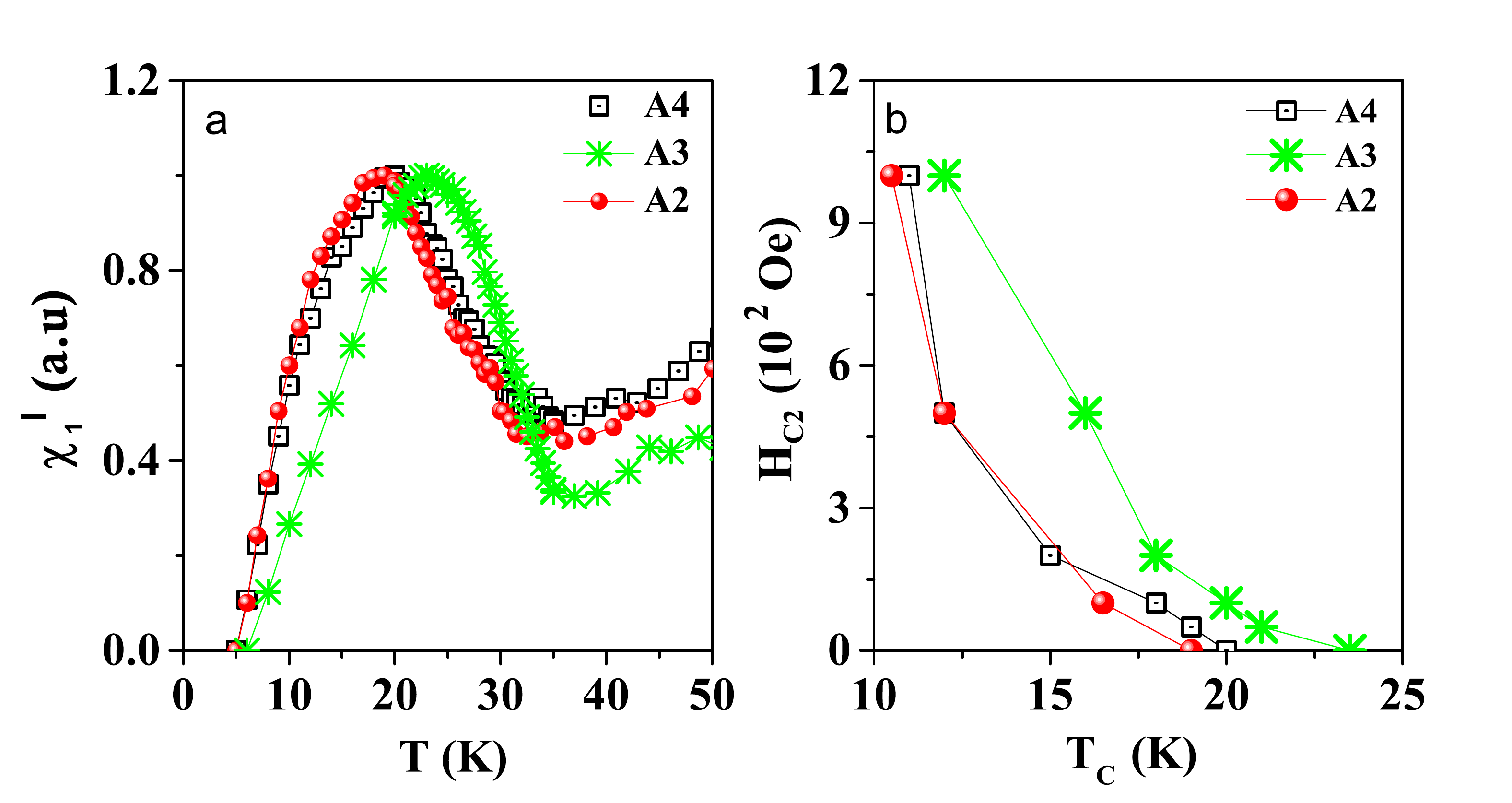}
		\vspace*{-4mm}
		\caption{\textit{\small{(Color online) (a) $\chi_{1}^{I}$ is plotted against temperature for A2, A3 and A4. (b) $H_{C2}$ is plotted against temperature for A2,A3 and A4 composites.}}}
		\label{fig:fig9}
	\end{figure*}
			
	These preformed SPM type AFM clusters are also observed to affect the superconducting state of LCu. Hence, a comprehensive understanding on the evolution of superconductivity in these composites require information on the critical thermodynamic parameters such as critical field ($H_{C2}$)  and critical current density ($J_C$) of LCu. $H_{C2}$  and $J_C$ are affected because of disorder and while the system size becomes comparable to certain characteristics length scale, like coherence length ($\xi_{0}$) and London penetration depth($\lambda_{l}$) \cite{c4r68,c4r68a,c4r68b}. Here the imaginary part of $\chi_{1}$ ($\chi_{1}^I$ ) has been used as a probing tool to realize the nature of $H_{C2}$ and $J_C$.  In ac-susceptibility measurement the phase lag between the ac-driving field and corresponding magnetization results in $\chi_{1}^I$. In case of a type-2 superconductor $\chi_{1}^I$ appears around $T_{S(onset)}$   because of vortex formation and pinning of it \cite{c4r65,c4r66,c4r67}. The maximum point of $\chi_{1}^I$ represents the temperature ($T_S$) at which the bulk superconductivity vanishes i.e. the critical current density ($J_C$) $\longrightarrow$0 and the applied magnetic field fully penetrates inside the superconductor \cite{c4r27,c4r65,c4r66,c4r67}. According to Bean model \cite{c4r27} the hysteresis loss (W) is inversely proportional to the critical current density ($J_C$), hence, at T\,$\longrightarrow$\,$T_S$, $J_C$$\longrightarrow$0, so W diverges at $T_S$ , as a result of that at T=\,$T_S$ peak in $\chi_{1}^I$ is observed [35-37].  The dc-field dependent behavior of $\chi_{1}^I$  around T$_{S(onset)}$ describes the anomaly is because of superconductivity \cite{c4r27} (see Section.\,C in the Supplemental Material). The temperature dependent $\chi_{1}^I$ plot of A2, A3 and A4 are shown in FIG.\,9(a). The onset point of $\chi_{1}^I$ for A3, A4 and A2 are observed around 34\,K, 31\,K and 30\,K respectively, whereas the diverging temperature of $\chi_{1}^I$ is observed at $\sim$24.5\,K, $\sim$20\,K and $\sim$19\,K for A3, A4 and A2, respectively. The crystalline size of LCu is same for all composites (as previously discussed), therefore according to Bean model and H$_C$, T$_S$ and J$_C$ phase diagram of a superconductor \cite{c4r27,c4r28,c4r29} the maximum point of $\chi_{1}^I$ is decided by the strength of the critical current density. In case of A3 maximum point of $\chi_{1}^I$ is observed at higher temperature than A2 and A4, which clearly indicates that the strength of critical current density is highest in case of composite A3, and according to the obtained temperature values, the amplitude of critical current density of the three composites can be expressed in the following descending order i.e., $J_{C}(A3)>J_{C}(A4)>J_{C}(A2)$. Therefore, at any particular temperature below T$_{S(onset)}$ the amplitude of critical field required to destroy the superconductivity is also highest for the corresponding material having highest value of critical current density. The plot of $H_{C2}$ against $T_S$ for A3, A2 and A4 is shown in FIG.\,9(b), where $H_{C2}$ can also be written in the following descending order $H_{C2}(A3)>H_{C2}(A4)>H_{C2}(A2)$. Therefore, all these results are indicating that the composites having larger AFM volume fraction show smaller diamagnetic fraction (discussed previously in FIG.\,4) and largest values of $H_{C2}$ and J$_C$. This kind of unusual increase of the amplitude of critical thermodynamical parameters and decrease of diamagnetic fraction (shown in FIG.\,4(b)) is often observed in the Quantum Size regime of superconductors, known as Quantum size effect (QSE). The QSE model was proposed by DeGennes and Tinkham \cite{c4r68,c4r68a,c4r68b}, where the mathematical expression for critical field ($H_{C2}$), critical current density (J$_C$) and effective London penetration depth ($\lambda_{eff}$) in the quantum limit can be represented as \cite{c4r68,c4r68a} 
	
	\begin{equation}
		H_{C2}\sim\frac{\xi_{0}\lambda_{l}}{r^{1.5}}
	\end{equation}
	\begin{equation}
		J_{C}\sim\frac{1}{r^{3}}
	\end{equation}
	\begin{equation}
		\lambda_{eff}=\lambda_{l}(1+\frac{\xi_{0}}{r})^{0.5}
	\end{equation}
		
	$\xi_{0}=\frac{0.18\hbar v_f}{k_{B}T_{C}}$ is the intrinsic coherence length, $v_F$ Fermi velocity, $\lambda_{l}$ London penetration depth and {$\bf\,r$} is the average size of the finite size superconducting clusters. According to \texttt{Eqn.\,5} and \texttt{Eqn.\,6} in the finite size regime the values of critical parameters are decided by the size of superconducting volume fraction, hence larger amplitude of $H_{C2}$ (and J$_C$) in case of A3 compared to A2 and A4 indicates the superconducting volume fraction is smaller in A3 compared to A2 and A4. According to \texttt{Eqn.\,7} the effective London penetration depth also increases in the finite size region, which causes the reduction of effective diamagnetic fraction (or Meissner fraction) and hence smaller value of superconducting volume fraction have been observed in case of A3 compared to A4 and A5 (as shown in FIG.\,4(b)).  XRD results are already depicting that there is no change of crystalline volume of LCu in any of the composite compared to parent LCu. So, decrease of the superconducting volume fraction and increase of $H_{C2}$ and J$_C$ is not due to reduction of crystalline size. Therefore, the SPM type AFM clusters which are formed within the bulk of LCu reduces the superconducting volume fraction, as the microscopic spin structure of the corresponding SPM cluster are static in nature and can not support superconductivity. It is already known from previous studies that any kind of static magnetic structure is unfavorable for superconductivity in cuprate superconductors, because the superconducting state of a cuprate evolves after diluting the antiferromagnetic network and only the dynamic AFM fluctuation can sustain with superconducting ordering \cite{c4t2,c4r1,c4r2}. Therefore, these observations are suggesting that the whole superconducting volume is divided into two phase separated regions, one phase consists of isolated SPM type AFM clusters and another phase is the finite size superconducting clusters.\\		
	So, from all the above discussion it can be claimed that when AFM fluctuation is stabilized by the exchange magnetic field of LCo, the AFM order emerges and superconducting volume fraction suppresses. This clearly indicates that the AFM type fluctuation can be a mediator of superconductivity in cuprate superconductor indicating that it is taking part in Cooper pairing.
	\vspace*{-3mm}
\section{Conclusion}
\vspace*{-3mm}
Composite of a superconductor (LCu) and a ferromagnet (LCo) has been prepared by solid state reaction method. In ac-susceptibility measurement an anomaly is observed above $T_{S(onset)}$ and it has been proved as the interface effect by changing the effective interface between LCu and LCo. It is proved through the studies of the linear ($\chi_{1}$) and nonlinear ac-susceptibilities ($\chi_{2}$, $\chi_{3}$) that the exchange field of LCo affects the copper spins and stabilizes the short range ordered AFM state within the bulk of LCu. These observations unambiguously indicates that the dynamic AFM fluctuation can be the mediator of copper pairing in cuprate superconductor substantiated by the observation of reduction of superconducting volume fraction when the AFM fluctuations are stabilized by formation of stable short range ordered SPM type AFM phase within the bulk. Further, the superconducting volume fraction reduce to the quantum size limit and two phase separated regions are formed, one is super-paramagnetic type AFM clusters and the rest is finite size superconducting volume. Another important inference is that, the reduced superconducting volume fraction shows quantum size effect (According to  DeGennes and Tinkham model), which is otherwise very difficult to achieve by preparing the nanoparticle of LCu, because of the structural limitation and degradation of oxygen stoichiometry. So these composites provide an unique path to understand the mechanism of Cooper pairing and also opens up a pathway to study the physical property of the finite size superconducting clusters of cuprate superconductors.
\vspace*{-2mm}
\section{Acknowledgment}
\vspace*{-2mm}
We are thankful to Dr. Mukul Gupta and Dr.N.P lalla for XRD measurements. Kranti Kumar Sharma and Dr.Santanu De are acknowledged for discussion and help during measurements. We are thankful to IRCC IIT Bombay MEMS department for TEM measurement.


\begin{thebibliography}{}
		\bibitem{c4t2}Chandra M. Varma, Rev. Mod. Phys. {\bf92}, 1 (2020).
	\bibitem{c4r1}D.J.Scalapino, Rev. Mod. Phys. {\bf84}, 1383 (2012).
	\bibitem{c4r2}M. P. M. Dean, G. Dellea, R. S. Springell, F. Yakhou-Harris, K. Kummer, N. B. Brookes, X. Liu, Y-J. Sun, J. Strle, T. Schmitt, et al., Nature Mater {\bf12}, 1019 (2013).
	\bibitem{c4t} K. Yamada, C. H. Lee, K. Kurahashi, J. Wada, S. Wakimoto, S. Ueki, H. Kimura, Y. Endoh, S. Hosoya, G. Shirane, R. J. Birgeneau, M. Greven, M. A. Kastner, and Y. J. Kim, Phys. Rev. B {\bf57}, 6165 (1998).
	\bibitem{c4t1}A. V. Balatsky, and P. Bourges, Phys. Rev. Lett {\bf82}, 5337 (1999).
	\bibitem{c4t3}S. Wakimoto, K. Yamada, J. M. Tranquada, C. D. Frost, R. J. Birgeneau, and H. Zhang, Phys. Rev.Lett {\bf98}, 247003 (2007).
	\bibitem{c4m1}G Aeppli , TE Mason, SM Hayden, HA Mook, J Kulda,  Science {\bf287}, 1432 (1997). 
		\bibitem{c4r50}J. M. Tranquada, B. J. Sternlieb, J. D. Axe, Y. Nakamura and S. Uchida, Nature London {\bf375}, 561 (1995). 
	\bibitem{c4r51} J. M. Tranquada, J. D. Axe, N. Ichikawa, Y. Nakamura, S. Uchida, and B. Nachumi,Phys. Rev. B {\bf54}, 7489 (1996). 
	\bibitem{c4r52}J. M. Tranquada, J. D. Axe, N. Ichikawa, A. R. Moodenbaugh, Y. Nakamura, and S. Uchida, Phys. Rev. Lett. {\bf78}, 338 (1997). 
		\bibitem{c4m2}V. A. Vas'ko, V. A. Larkin, P. A. Kraus, K. R. Nikolaev, D. E. Grupp, C. A. Nordman, and A. M. Goldman, Phys. Rev. Lett. {\bf78}, 1134 (1997).
	\bibitem{c4m3}V. Peña, Z. Sefrioui, D. Arias, C. Leon, J. Santamaria, J. L. Martinez, S. G. E. te Velthuis, and A. Hoffmann, Phys. Rev. Lett.{\bf94}, 057002 (2005). 
	\bibitem{c4r9}V. Peña, Z. Sefrioui, D. Arias, C. Leon, J. Santamaria, M. Varela, S. J. Pennycook, and J. L. Martinez, Phys. Rev.b. {\bf69}, 224502 (2004). 
	\bibitem{c4r10} J. Chakhalian, J. W. Freeland, G. Srajer, J. Strempfer, G. Khaliullin, J. C. Cezar, T. Charlton, R. Dalgliesh, C. Bernhard, G. Cristiani, H.-U. Habermeier and B. Keimer, Nature Phys.{\bf2}, 244 (2006). 
	\bibitem{c4r11} D. K. Satapathy, M. A. Uribe-Laverde, I. Marozau, V. K. Malik, S. Das, Th. Wagner, C. Marcelot, J. Stahn, S. Brück, A. Rühm, S. Macke, T. Tietze, E. Goering, A. Frañó, J. -H. Kim, M. Wu, E. Benckiser, B. Keimer, A. Devishvili, B. P. Toperverg, M. Merz, P. Nagel, S. Schuppler, and C. Bernhard,  Phys. Rev.Lett {\bf108}, 197201 (2012). 
	
	\bibitem{c4r13} N. Driza, S. Blanco-Canosa, M. Bakr, S. Soltan, M. Khalid, L. Mustafa, K. Kawashima, G. Christiani, H-U. Habermeier, G. Khaliullin, C. Ulrich, M. Le Tacon and B. Keimer, Nature Mater.{\bf11}, 675 (2012). 
	\bibitem{c4r14}A. Frano, S. Blanco-Canosa, E. Schierle, Y. Lu, M. Wu, M. Bluschke, M. Minola, G. Christiani, H. U. Habermeier, G. Logvenov, Y. Wang, P. A. van Aken, E. Benckiser, E. Weschke, M. Le Tacon and B. Keimer, Nature Mater. {\bf15}, 831 (2016).
	
	\bibitem{c4r15}Junfeng He, Padraic Shafer, Thomas R. Mion, Vu Thanh Tra, Qing He, J. Kong, Y.-D. Chuang, W. L. Yang, M. J. Graf, J.-Y. Lin, Y.-H. Chu, E. Arenholz and Rui-Hua He, Nature Commun. {\bf7}, 10852 (2016). 
		\bibitem{c4r16}J. Chakhalian, J. W. Freeland, H.-U. Habermeier, G. Cristiani,
		G. Khaliullin, M. van Veenendaal, B. Keimer, Science {\bf318}, 1114 (2007).  
		
	\bibitem{c4y1}Keyan Li and Dongfeng Xue, J. Phys. Chem. A, {\bf110}, 11332 (2006). 
	
	\bibitem{c4y2}H. Y. Hwang, Y. Iwasa, M. Kawasaki, B. Keimer, N. Nagaosa and Y. Tokura , Nature Mater.{\bf11}, 103 (2012). 
	\bibitem{c4y3}Jaewoo Jeong, Nagaphani Aetukuri, Tanja Graf, Thomas D. Schladt,	Mahesh G. Samant, Stuart S. P. Parkin, Science {\bf339}, 1402 (2013).

		\bibitem{c4a1}S. Komori, A. Di Bernardo, A.I.Buzdin, M.G. Blamier, and J.W.A. Robinson, Phys. Rev.Lett {\bf 121}, 077003 (2018).
		 I. Bozovic, G. Logvenov, I. Belca, B. Narimbetov, and I. Sveklo, Phys. Rev. Lett. {\bf89}, 107001 (2002).
		 J.P. Locquet, J. Perret, J. Fompeyrine, E. Machler, J. W. Seo and G. Van Tendeloo , Nature {\bf394}, 453 (1998).
	\bibitem{c4r19} Daniel Hsu, T. Geetha Kumary, L. Lin, and J. G. Lin, Phys. Rev. B {\bf74}, 214504 (2006). 
		\bibitem{c4r17} N. Biškup, S. Das, J. M. Gonzalez-Calbet, C. Bernhard, and M. Varela, Phys. Rev. B {\bf91}, 205132 (2015). 
	\bibitem{c4r18} S. Das, K. Sen, I. Marozau, M. A. Uribe-Laverde, N. Biskup, M. Varela, Y. Khaydukov, O. Soltwedel, T. Keller, M. Döbeli, C. W. Schneider, and C. Bernhard, Phys. Rev. B {\bf89}, 094511 (2014).
		\bibitem{c4r20} Masayuki Itoh , Ikuomi Natori , Satoshi Kubota , Kiyoichiro Motoya, J. Magn. Magn.Mater. {\bf140}, 1811 (1995).
	\bibitem{c4r21} J. Wu and C. Leighton, Phys. Rev. B  {\bf67}, 174408 (2003). 
	\bibitem{c4r22} P S Anil Kumar, P A Joy and S K Date, J. Phys.Condens. Matter {\bf10}, L487 (1998). 
	\bibitem{c4r23}Vivek K. Malik, Chi Hieu Vo, Elke Arenholz, Andreas Scholl, Anthony T. Young, and Yayoi Takamura, J. Appl. Phys. {\bf113}, 153907 (2013).
	
		\bibitem{c4r58}C. Panagopoulos, B.D. Rainford, J.R. Cooper, C.A. Scott Physica C {\bf341}, 843 (2000). 
	\bibitem{c4r59} M.-H. Julien, A. Campana, A. Rigamonti, P. Carretta, F. Borsa, P. Kuhns, A. P. Reyes, W. G. Moulton, M. Horvatić, C. Berthier, A. Vietkin, and A. Revcolevschi, Phys. Rev. B {\bf63}, 144508, (2001). 
	\bibitem{c4r60} C. Panagopoulos, J. L. Tallon, B. D. Rainford, T. Xiang, J. R. Cooper, and C. A. Scott, Phys. Rev. B {\bf66}, 064501 (2002). 
		\bibitem{c4r24}Biswajit Dutta, Kranti Kumar, and A. Banerjee, AIP Conference Proceedings {\bf2115}, 030515 (2019). 
			\bibitem{c4r27} Charles P. Bean., Rev. Mod. Phys. {\bf36}, 31 (1964). 
		\bibitem{c4r28} P.W. Anderson and Y.B. Kim, Rev. Mod. Phys. {\bf36}, 39 (1964). 
		\bibitem{c4r29} G. Ravi Kumar and P. Chaddah, Phys. Rev. B  {\bf39}, 4704 (1989). 
		\bibitem{c4r30} K.H.Müller, J.C.Macfarlane, R.Driver, Physica C {\bf158}, 366 (1989).
			\bibitem{c4x}D. Bhattacharya, L. C. Pathak, S. K. Mishra, D. Sen and K. L. Chopra, Appl. Phys. Lett. {\bf57}, 2145 (1990).
				\bibitem{c4x1}A. Bajpai and A. Banerjee, Phys. Rev. B {\bf55}, 12 439 (1997).
			\bibitem{c4x2}A. Bajpai and A. Banerjee, Phys. Rev. B {\bf62}, 8996 (2000).
				\bibitem{c4x3}S. Mukherjee and R. Ranganathan, Phys. Rev. B {\bf54}, 9267 (1996).
			\bibitem{c4x4} Toshikaju Sato and Yoshihito Miyako, J. Phys. Soc. Jpn. {\bf51}, 1394 (1981).
			\bibitem{c4x5} S.Fujiki and S.Katsura, Prog.Theo.Phys., {\bf65}, 1130 (1981).
			
					\bibitem{r50}J. M. Tranquada, B. J. Sternlieb, J. D. Axe, Y. Nakamura and S. Uchida , Nature London {\bf375}, 561 (1995).
							
						\bibitem{r39} Biswajit Dutta, Kranti Kumar, N. Ghodke, and  A. Banerjee, Rev. Sci. Instrum. {\bf91}, 123905 (2020). A Bajpai, A Banerjee,  Rev. Sci. Instrum. {\bf 68}, 4075 (1997).
	\bibitem{r39a}T. Bitoh, K. Ohba, M. Takamatsu, T. Shirane, and S. Chikazawa, J. Magn.Magn. Mater. {\bf154}, 59 (1996). S. Mukherjee, R. Ranganathan, S.B. Roy, Solid State Commun. {\bf98}, 321(1996). A. Chakravarti and R. Ranganathan, Solid State Commun. 82, 591 (1992).
				\bibitem{x4} Toshikaju Sato and Yoshihito Miyako, J. Phys. Soc. Jpn. {\bf51}, 1394 (1981).
				\bibitem{x6} G.Sinha and A.Majumdar,J.Magn.Magn.Mater.{\bf185}, 18 (1998).
				\bibitem{x7}A.Chakravarti, R.Ranganathan, C.Bansal, Solid State Commun. {\bf82}, 591 (1992).
			\bibitem{c4r41} K.Binder and A.P.Young, Rev. of Mod. Phys., {\bf58}, 807 (1986).
				\bibitem{c4r47} A. Bajpai and A. Banerjee, J. Phys. Condens. Matter {\bf13}, 637 (2001). A. K. Pramanik and A. Banerjee, Phys. Rev. b. {\bf82}, 094402 (2010).
			\bibitem{c4r48} Sunil Nair and A. Banerjee, Phys. Rev. Lett. {\bf93}, 117204 (2004).
				\bibitem{c4r48a} Sunil Nair and A. Banerjee, Phys. Rev. b. {\bf68}, 094408, (2003).
					\bibitem{c4r42} Masuo Suzuki, Progress of Theoretical Physics, {\bf58}, 1151, (1977).
					\bibitem{c4r44} E.P.Wohlfarth Phys.Lett. {\bf70A}, 489 (1979).
				\bibitem{c4r45}T.Bitoh, K.Ohba, M.Takamatsu, T.Shirane, S.Chikazawa, J. Magn. Magn. Mater. {\bf154}, 59 (1996).
				\bibitem{c4r53} A. J. Achkar, R. Sutarto, X. Mao, F. He, A. Frano, S. Blanco-Canosa, M. Le Tacon, G. Ghiringhelli, L. Braicovich, M. Minola, M. Moretti Sala, C. Mazzoli, Ruixing Liang, D. A. Bonn, W. N. Hardy, B. Keimer, G. A. Sawatzky, and D. G. Hawthorn, Phys. Rev. Lett. {\bf109}, 167001 (2012).
			\bibitem{c4r54} A.J. Achkar, X.Mao, Christopher McMahon, R.Sutarto, F.He, Ruixing Liang, D.A. Bonn, W.N.Hardy, and D.G.Hawthorn, Phys. Rev. Lett. {\bf113}, 107002 (2014).
			\bibitem{c4r55} S. Sachdev, E. Demler, Phys. Rev. B {\bf69}, 144504 (2004).
			\bibitem{c4r56}Lauren E. Hayward, David G. Hawthorn, Roger G. Melko, Subir Sachdev, Science {\bf343},1336 (2014).
			
		\bibitem{c4r61}W. H. Meiklejohn and C. P. Bean, Phys. Rev. {\bf102}, 1413 (1956).
		\bibitem{c4r62}D. Mauri, H. C. Siegmann, P. S. Bagus, and E. Kay, J. Appl. Phys.{\bf62} , 3047 (1987).
		\bibitem {c4r63} N. C. Koon.,Phys. Rev. Lett. {\bf78}, 4865 (1997).
		\bibitem{c4r64} Shilpi Karmakar, S. Taran, Esa Bose, B. K. Chaudhuri, C. P. Sun, C. L. Huang, and H. D. Yang,  Phys. Rev. B {\bf77}, 144409 (2008).
		
	\bibitem{c4r68}Sangita Bose and Pushan Ayyub, Rep. Prog. Phys. {\bf77}, 116503 (2014).
	\bibitem{c4r68a}P.G.De Gennes  and M. Tinkham , Physics {\bf1}, 107(1964).
	\bibitem{c4r68b}M.Tinkham, Phys. Rev. {\bf110}, 26 (1958).
		\bibitem{c4r65}L. Civale, T.K. Worthington, L. Krusin-Elbaum and F. Holtzberg, in Magnetic Susceptibility of Superconductors
	and Other Spin Systems, edited by R. A. Hein, T. L. Francavilla, and
	D. H. Liebenberg (Plenum, New York, 1991), pp 313-332, and references therein.
	\bibitem{c4r66}Xinsheng Ling, and Joseph I. Budnick, in Magnetic Susceptibility of Superconductors and Other Spin Systems, edited by R. A. Hein, T. L. Francavilla, and
	D. H. Liebenberg (Plenum, New York, 1991), pp 377-388, and references therein.
	\bibitem{c4r67} S. Ramakrishnan, Ravi Kumar, C. V. Tomy, A. K. Grover, S. K. Malik, P. Chaddah, in Magnetic Susceptibility of Superconductors
	and Other Spin Systems, edited by R. A. Hein, T. L. Francavilla, and
	D. H. Liebenberg (Plenum, New York, 1991), pp 389-404, and references therein.
	
		
		
	\end{thebibliography}
\end{document}


\title{Supplemental Material:
	\vspace{4.5mm}
	\\ Ferromagnetic exchange field stabilized antiferromagnetic ordering in a cuprate superconductor} 
	\author{Biswajit Dutta and A. Banerjee}
	\affiliation{UGC-DAE Consortium for Scientific Research, University Campus, Khandwa Road, Indore-452001, India.}
	
	\maketitle
		\subsection{Transmission electron data analysis}
	\vspace{-4.5mm}
		The microstructures of the composite materials have been studied by high-resolution transmission electron microscopy (HRTEM, Thermo Scientific, Themis 300 G3, 300 kV). Fig.\,1 shows the Scanning Transmission electron microscopy-high-angle annular dark-field (STEM HAADF) images and the corresponding elemental images of La, Sr, O, Co, Cu as well as C for the composite A3. It can be seen that all elements are uniformly dispersed in the composite A3, an indirect indication of the formation of a face-to-face contact interface between the superconductor $La_{1.85}Sr_{0.15}CuO_{4}$ (LCu) and the ferromagnet $La_{0.6}Sr_{0.4}CoO_{3}$ (LCo).  Fig.\,2(a) shows the typical TEM image of A3 composite with irregular shape crystalline pallets. The microstructure of the interface between LCu and LCo crystallites have been investigated by high-resolution Transmission electron microscopy (HRTEM). Various areas in the composite are spotlighted to record the high-resolution image of the interface. Fig.\,2(a) shows the normal TEM image of A3, and Fig.\, 2(b)- Fig.\,2(d) show the HRTEM images of the distinctly connected LCu and LCo crystallites.	The fringes of the marked lattice spacing of 0.38 nm in Fig.\,2(d) correspond to the lattice parameter of LCu. Importantly, the contact interface between  LCu and LCo is very sharp. There is no sign of other crystal geometry or crystallites revealed across the interface. These images are depicting very sharp interfaces between LCu and LCo, and also ruled out the possibilities of intermixing or chemical reaction. The expanded view of Fig.\,2(c) is shown in Fig.\,3, the darkest region corresponds to LCo whereas the lighter portion corresponds to LCu. The inter-planner spacing of LCo is calculated around 0.5\,nm while for LCu it is around 0.38\,nm, these values are closely matched with the a and b axis lattice parameters of LCo and LCu obtained from the XRD results (mentioned in {\bf TABLE\,II} of the main manuscript). 
		\begin{figure*}[t]
		\centering
		\includegraphics[width=16cm]{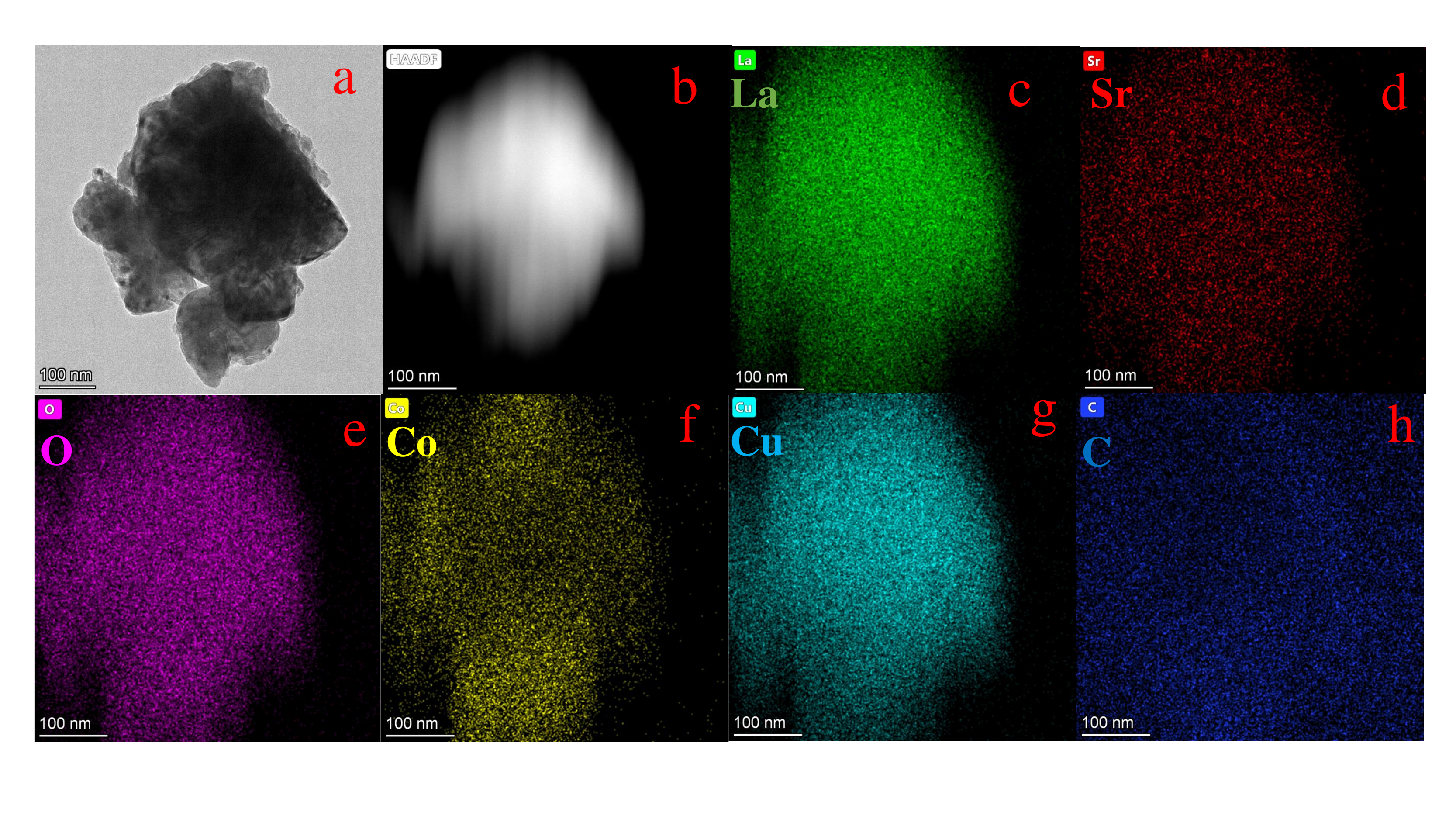}
		\caption{\textit{\small{(Color online) (a) Elemental mapping images of A3 composite by STEM-HAADF. (b) STEM-HAADF image, (b) combined La/Sr/O/Co/Cu/C chemical mapping image and the distribution of (c) Sr, (d) O (e) Co (f) Cu (g) and (C) elements on A3 composite.}}}
		\label{fig:figS1}
	\end{figure*}
	
	\begin{figure}[htbp]
		\centering
		\includegraphics[width=16cm]{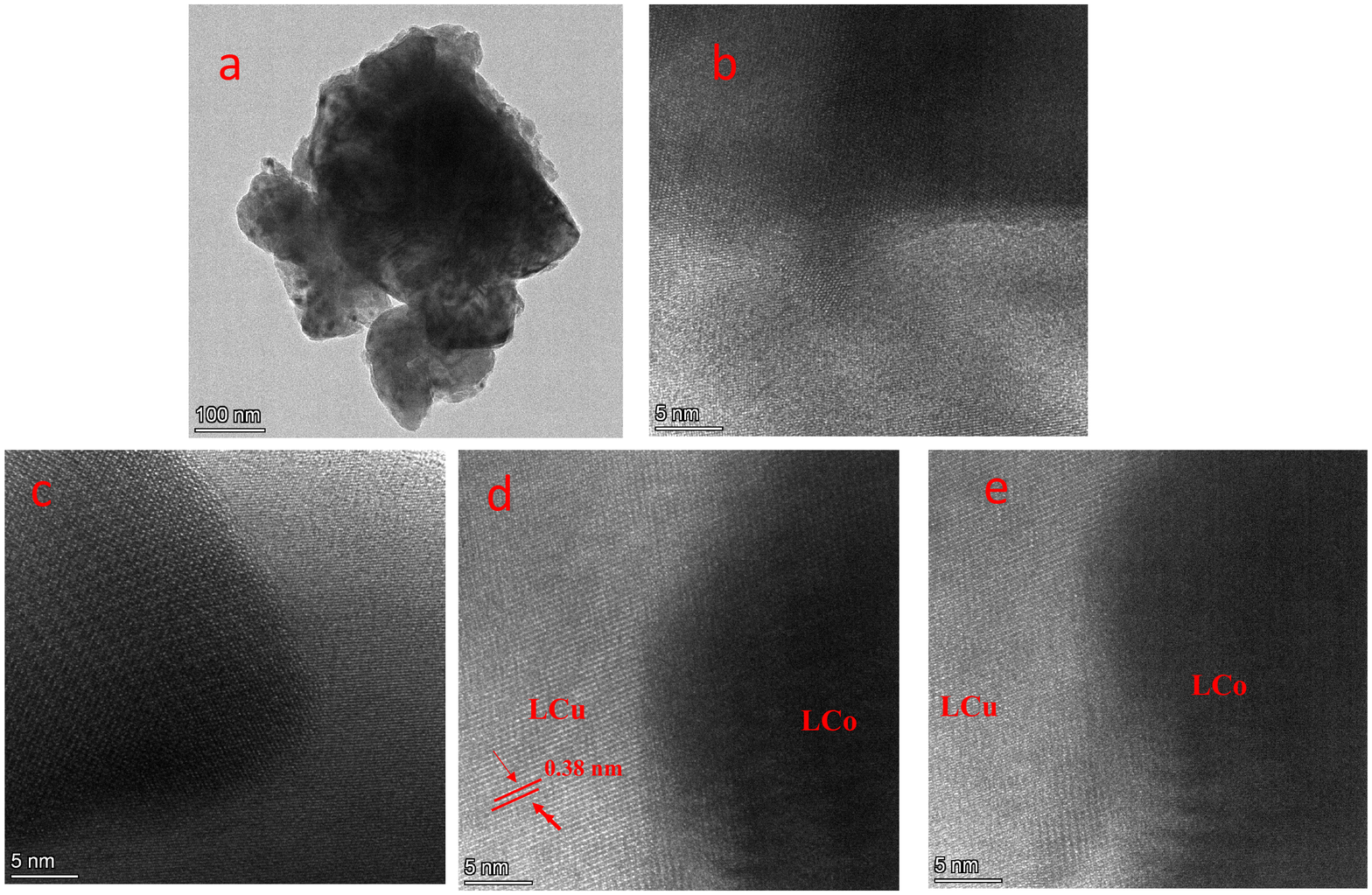}
		\caption{\textit{\small{(Color online) (a) TEM image (b),(c),(d) and (e) are high resolution TEM  images of composite A3.}}}
		\label{fig:fig2}
	\end{figure}
	
	\begin{figure}[t]
		\centering
		\includegraphics[width=16cm]{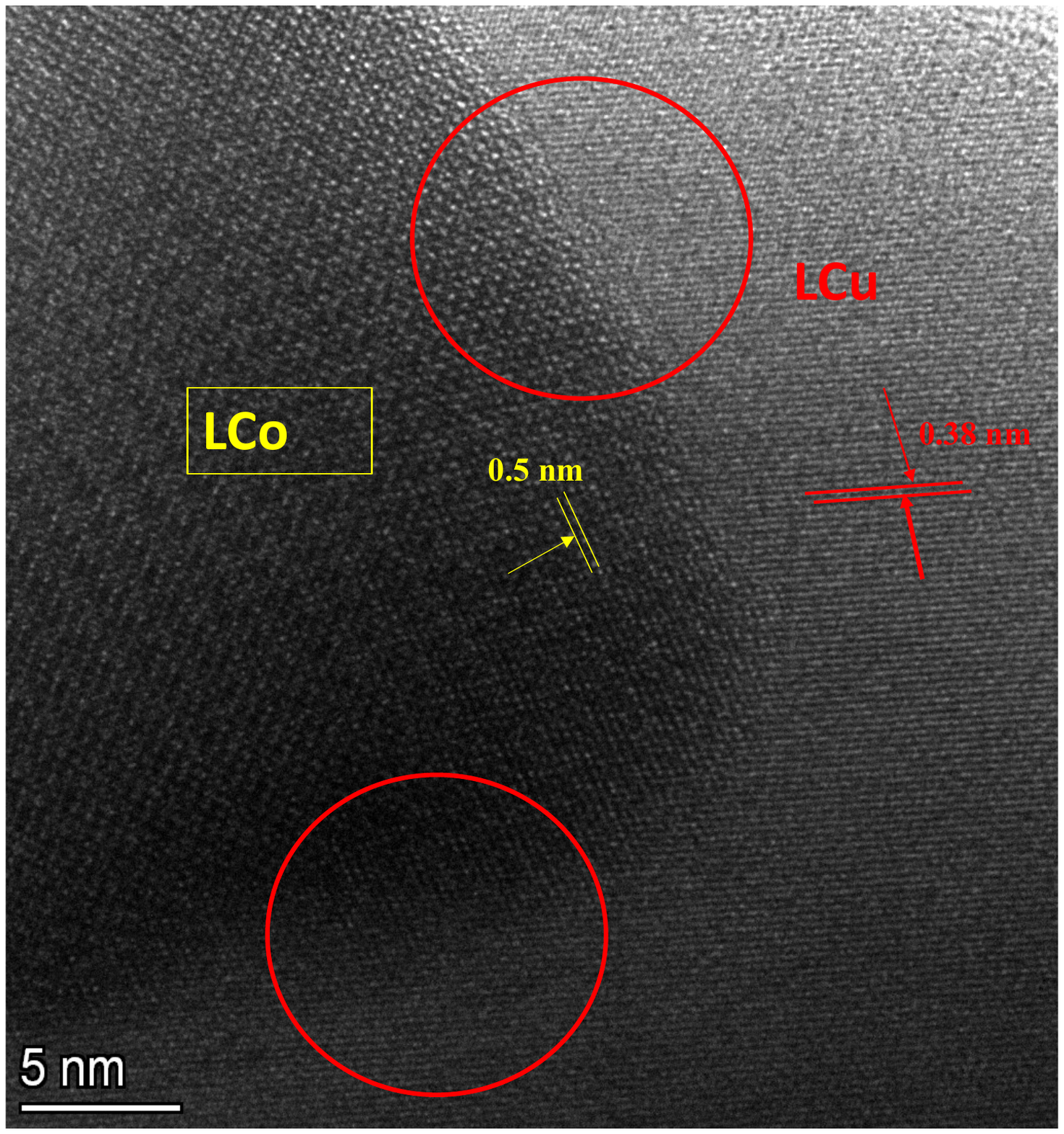}
		\caption{\textit{\small{(Color online) HRTEM image  of composite A3.}}}
		\label{fig:fig3}
	\end{figure}

	\begin{figure}[t]
		\centering
		\includegraphics[width=16cm]{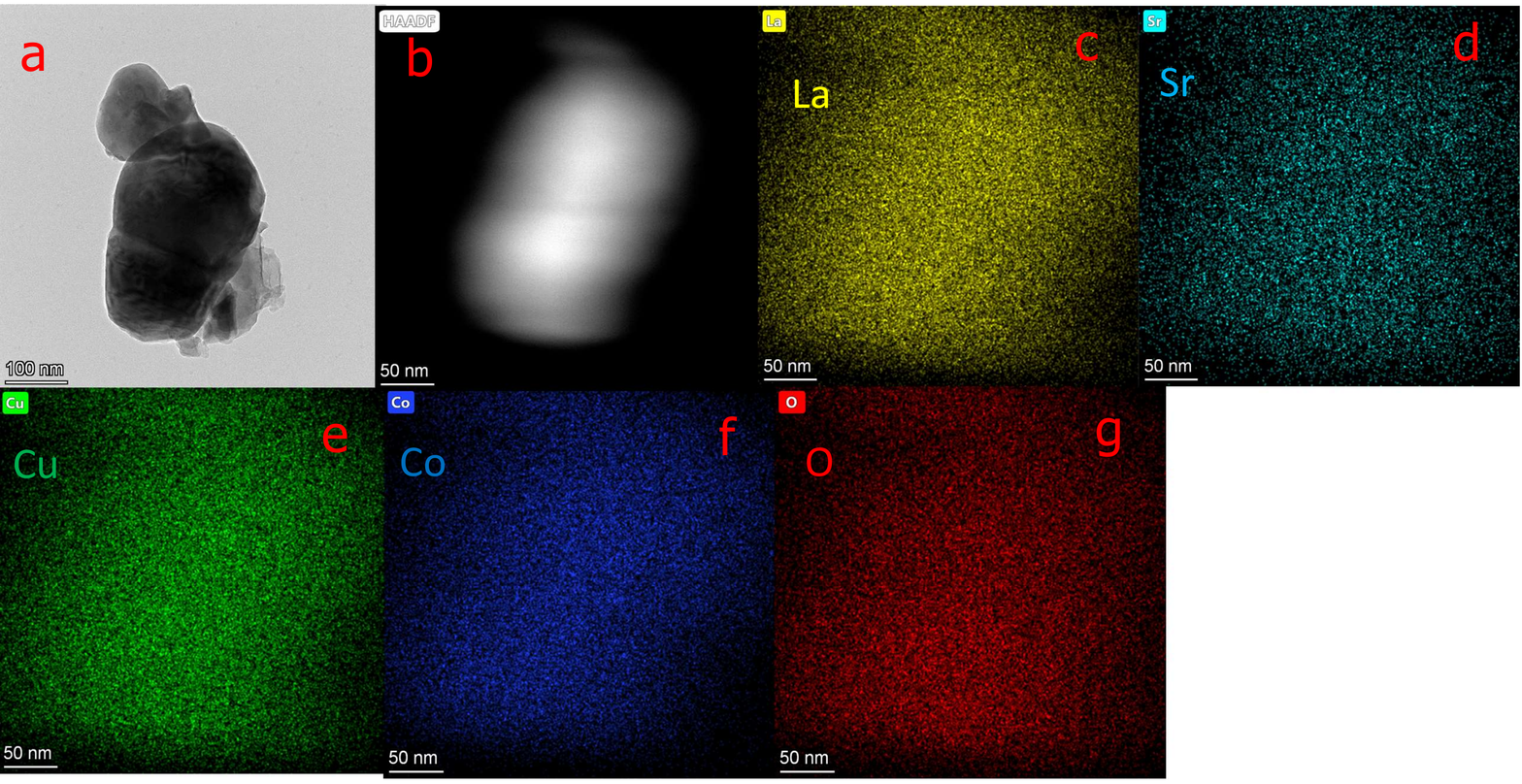}
		\caption{\textit{\small{(Color online) (a) Elemental mapping images of A2 composite by STEM-HAADF. (b) STEM-HAADF image, (b) combined La/Sr/Cu/Co/O chemical mapping image and the distribution of (c) La, (d) Sr (e) Cu (f) Co (g) and O elements on A2 composite.}}}
		\label{fig:fig4}
	\end{figure}
	
	\begin{figure}[htbp]
		\centering
		\includegraphics[width=16cm]{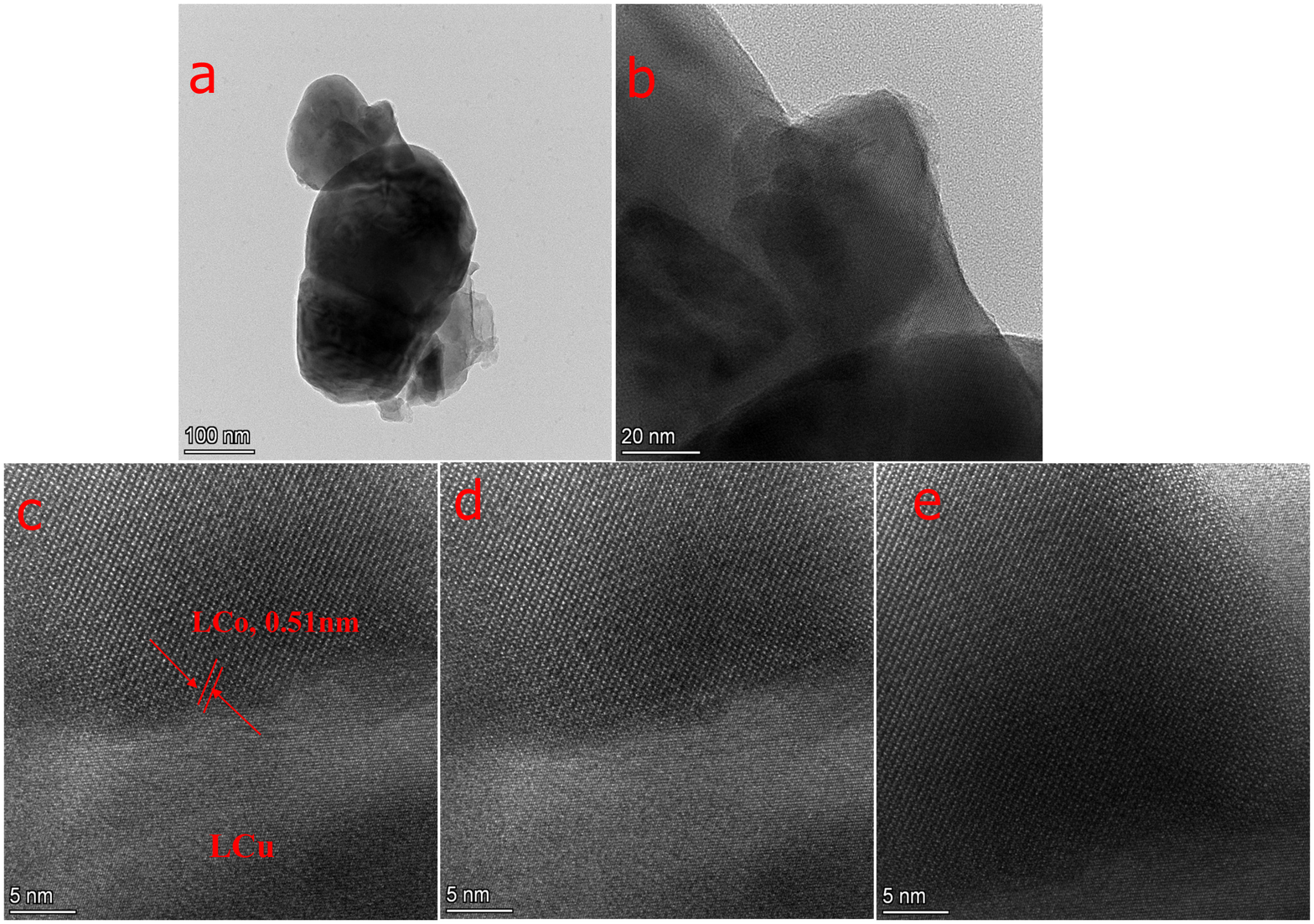}
		\caption{\textit{\small{(Color online) (a) TEM image, (b),(c),(d), and (e) are high resolution TEM  images of composite A2.}}}
		\label{fig:fig5}
	\end{figure}
	
	The STEM-HAADF image of the composite (A2) is shown in Fig.\,4. It also shows the uniform distribution of all elements like La, Sr, Cu, Co, and O shown in Fig.\,4(c), Fig.\,4(d), Fig.\,4(e), Fig.\,4(f), and Fig.\,4(g), respectively. which is an indirect sign of face-to-face connection between LCu and LCo domains. The TEM (Transmission Electron Microscopy) image of A2 is shown in Fig.\,5. HRTEM (High Resolution Transmission Electron Microscopy) images are shown in Fig.\,5(c), Fig.\,5(d), Fig.\,5(e) at various resolution. A clear phase boundary is seen between LCu and LCo grain. There are no other phases observed across the phase boundary of LCu and LCo. 
	
	Therefore, the TEM (Transmission Electron Microscopy) images of the composite A2 and A3 strongly suggest the fact that the interface between LCu and LCo is very sharp and depicts the absence of any foreign phase across the interface of the composite.

\subsection{Minor loop study of the composite}
			\begin{figure}[t]
			\centering
			\includegraphics[width=9cm]{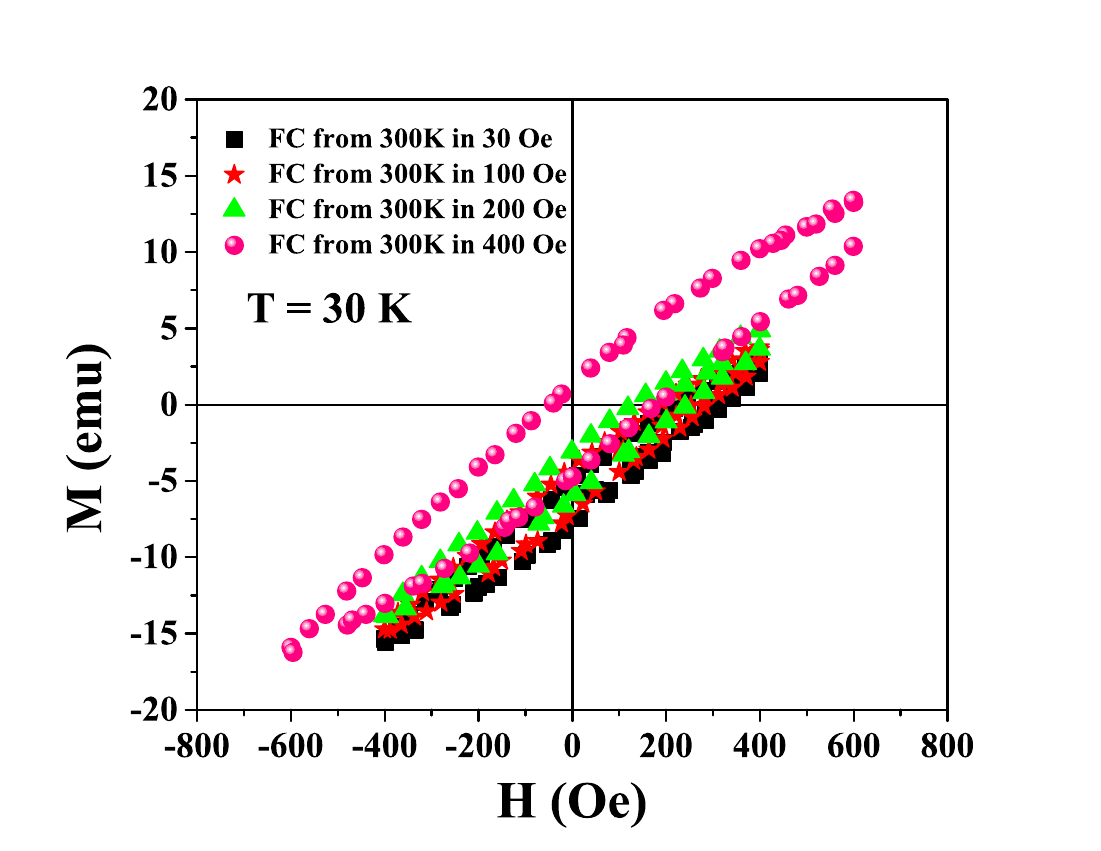}
			\caption{\textit{\small{(Color online) (a) The magnetic isotherm of composite A3 at 30\,K at cooling conditions of 30\,Oe, 100\,Oe, 200\,Oe, and 400 Oe.}}}
			\label{fig:fig6}
				\end{figure}
The  field cool magnetic isotherm measurement of the composite A3 at 30\,K is shown in FIG.\,6. The following protocol is followed to measure the magnetic hysteresis loop. The sample is initially cooled to 30\,Oe (the black square) from room temperature, and the magnetic isotherm measurements are then taken between $\pm$400\,Oe. The obtained curve exhibits a minor loop behaviour as well as a sizable amount of loop asymmetries, which is an indication of exchange bias.  similarly the field cool isotherm measurement is done in other field cooling conditions like 100,200 and 400\,Oe . As the cooling field is increased, the bulk magnetization behaviour of the ferromagnet (i.e., $La_{0.6}Sr_{0.4}CoO_{3}$ (LCo) ) begins to dominate, which causes the exchange bias effect to decrease. It  unambiguously proves that AFM clusters have formed within LCu and their proximity to FM LCo might give rise to asymmetry in the magnetic isotherm of the composite (shown in FIG.\,6).	
	\subsection{Dc field dependent $\chi_{1}^I$ }
\begin{figure}[t]
	\centering
	\includegraphics[width=9cm]{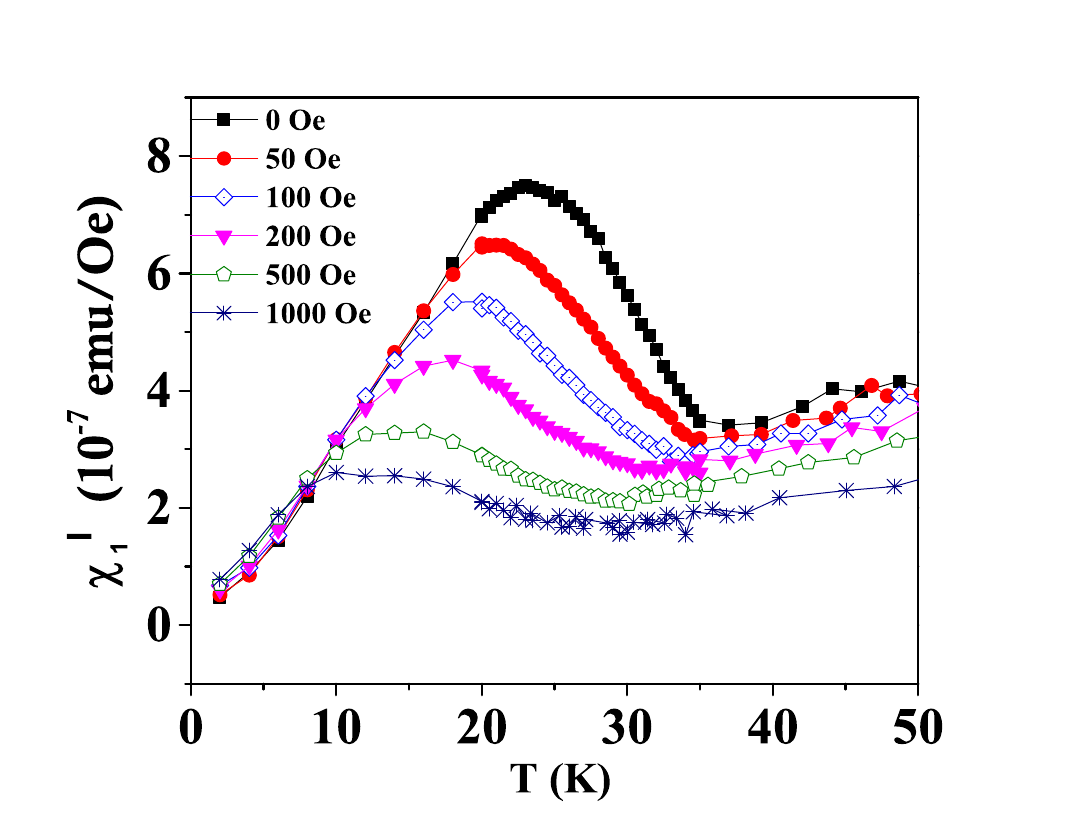}
	\caption{\textit{\small{(Color online) (a) $\chi_{1}^I$ plot of composite A3 at 0, 50, 100, 200, 500 and 1000 Oe superimposed dc field.}}}
	\label{fig:fig7}
\end{figure}
The onset temperature and the peak temperature of $\chi_ {1} ^I$ decrease with increasing the amplitude of the dc bias field shown in FIG.\,7, which is a typical signature of superconductor [Ref. 35] i.e., decrement of superconductivity while increasing the amplitude of dc bias field.